\documentclass[traditabstract,times]{aa} 

\newcommand{\ergps}{erg\thinspace s$^{-1}$}
\newcommand{\ergpspsqcm}{erg\thinspace s$^{-1}$\thinspace cm$^{-2}$}
\newcommand{\psqcm}{cm$^{-2}$}
\newcommand{\nH}{$N_{\rm H}$}

\usepackage{graphicx}
\begin{document}
\title{Fe~K emission from active galaxies in the COSMOS field}


   \author{K. Iwasawa\inst{1}\thanks{Email: kazushi.iwasawa@icc.ub.edu}
          \and
          V. Mainieri\inst{2}
          \and
M. Brusa\inst{3}
\and
A. Comastri\inst{4}
\and
R. Gilli\inst{4}
\and
C. Vignali\inst{5}
\and
G. Hasinger\inst{6}
\and
D.B. Sanders\inst{6}
\and
N.~Cappelluti\inst{4}
\and
C.D. Impey\inst{7,8}
\and
A. Koekemoer\inst{9}
\and
G. Lanzuisi\inst{10}
\and
E. Lusso\inst{5,11}
\and
A. Merloni\inst{3}
\and
M. Salvato\inst{12}
\and
Y.~Taniguchi\inst{13}
\and
J.R.~Trump\inst{14}
}
\institute{ICREA and Institut de Ci\`encies del Cosmos (ICC), Universitat de Barcelona (IEEC-UB), Mart\'i i Franqu\`es, 1, 08028 Barcelona, Spain
         \and
European Southern Observatory, Karl-Schwarzschild-Stra\ss e 2, 85748 Garching, Germany
            \and
Max Planck Institut f\"ur extraterrestrische Physik, Gie\ss enbachstra\ss e, 85748 Garching, Germany
\and            
INAF - Osservatorio Astronomico di Bologna, Via Ranzani, 1, 40127 Bologna, Italy
\and
Universit\`a di Bologna - Dipartimento di Astronomia, Via Ranzani, 1, 40127 Bologna, Italy
\and
Institute for Astronomy, University of Hawaii, 2680 Woodlawn Drive, Honolulu, Hawaii 96822-1839, USA
\and
Department of Astrophysical Science, University of Princeton, Peyton Hall 103, Princeton NJ 08544, USA
\and
Steward Observatory, University of Arizona, 933 North Cherry Avenue, Tucson AZ 85721, USA
\and
Space Telescope Science Institute, 3700 San Martin Drive, Baltimore MD 21218, USA
\and
INAF - IASF Roma, Via Fosso del Cavaliere 100, 00133 Roma, Italy
\and
Max-Planck-Institut f\"ur Astronomie, K\"onigstuhl 17, 69117 Heidelberg, Germany
\and
Max-Planck-Institut f\"ur Plasmaphysik, Boltzmannstra\ss e 2, 85748 Garching, Germany
\and
Research Center for Space and Cosmic Evolution, Ehime University, 2-5 Bunkyo-cho, 790-8577 Matsuyama, Japan
\and
University of California Observatories/Lick Observatory, University of California, Santa Cruz CA 95064, USA
          }

   \date{Accepted on 8 November 2011}

 
   \abstract{We present a rest-frame spectral stacking analysis of $\sim
     1000$ X-ray sources detected in the XMM-COSMOS field in order to investigate
     the iron K line properties of active galaxies beyond redshift $z\sim
     1$. In Type I AGN that have a typical X-ray luminosity of $L_{\rm X}\sim 1.5\times 10^{44}$ (\ergps) and $z\sim 1.6$, the cold Fe~K at
     6.4 keV is weak ($EW \sim 0.05$~keV), in agreement with the known
     trend. In contrast, high-ionization lines of Fe {\sc xxv} and Fe
     {\sc xxvi} are pronounced. These high-ionization Fe~K lines appear to
     have a connection with high accretion rates. While no broad Fe
     emission is detected in the total spectrum, it might be present, 
     albeit at low significance ($\sim 2\sigma $), when the X-ray
     luminosity is restricted to the range below $3\times 10^{44}$ \ergps, or
     when an intermediate range of Eddington ratio around $\lambda\sim 0.1$
     is selected. In Type II AGN, both cold and high-ionzation lines
     become weak with increasing X-ray luminosity. However, strong
     high-ionization Fe~K ($EW \sim 0.3$ keV) is detected in the
     spectrum of objects at $z>2$, while no 6.4 keV line is found. It
     is then found that the primary source of the high-ionization Fe~K
     emission is those objects detected with Spitzer-MIPS at 24$\mu$m. Given their 
     median redshift of $z\simeq 2.5$, their bolometric luminosity is
     likely to reach $10^{13}L_{\odot}$ and the MIPS-detected emission
     most likely originates from hot dust heated by embedded AGN, probably
     accreting at high Eddington ratio. These properties match those of rapidly
     growing black holes in ultra-luminous infrared galaxies at the
     interesting epoch ($z\sim 2$-$3$) of galaxy formation.}

\titlerunning{Fe~K emission of active galaxies in COSMOS}
\authorrunning{K. Iwasawa et al.}
   \maketitle
%

\section{Introduction}

The iron K-shell (Fe~K) emission at 6-7 keV is the most prominent
spectral line in the X-ray spectra of active galaxies and serves as a useful probe
of various physical properties of galactic nuclei. However, given 
the current available instrumentation, X-ray observations of active galactic
nuclei (AGN) are still in the photon-limited regime in most cases, and
the study of Fe~K lines is limited to nearby bright objects. For
example, $\sim 85$ \% of the unobscured AGN sample of Bianchi et al
(CAIXA, 2009), which is the largest compilation of unobscured AGN in
the XMM-Newton archive, lie at redshift $z\leq 0.4$. Beyond $z =1$, the Fe~K 
line properties of active galaxies are virtually unexplored
apart from a few exceptions for which very
long exposure observations are available, e.g., Comastri et al (2011), Feruglio et al (2011), Norman et al (2002).

With the relatively large area of the sky ($\sim $2-deg$^2$) covered by
the COSMOS survey (Scoville et al 2007) with its depth down to faint
fluxes probed by XMM-Newton (Cappelluti et al 2009), a significant
number of luminous X-ray sources at $z\geq 1$, most of which are AGN,
are detected. While detection of a Fe~K line in individual sources is
limited to a very small fraction of the X-ray objects (Mainieri et al
2007), we have made use of the large XMM-COSMOS dataset (Hasinger et
al 2007) and spectral stacking to obtain integrated signals of the
faint Fe emission for understanding the physical properties of active
galaxies at high redshift.

Here we first summarize a few relevant facts concerning the Fe~K line, obtained from
observations of nearby AGN: 1) The emission feature is dominated by
the 6.4 keV line (up to Fe {\sc xvii}), indicating that a cold medium
illuminated by an AGN is the primary source of the line
emission. High-ionization lines from Fe {\sc xxv} at 6.7 keV and Fe
{\sc xxvi} at 6.97 keV are also seen in good quality data (e.g.,
Bianchi et al 2007) but are much weaker than the cold line; 2) In
unobscured AGN, a line profile with a redward extension is observed
in a significant fraction of nearby Seyfert galaxies (de la Calle
P\'erez et al 2010; Fukazawa et al 2011; Bianchi et al 2009; Nandra et
al 2007). Its origin has been under debate,  but it may be a result of
spectral distortion due to strong gravity, and therefore could serve as a
powerful probe of the accretion flow in the relativistic region in the
vicinity of a black hole (e.g., Fabian et al 2000); and 3) In  
heavily obscured AGN, observed X-rays are suppressed, and thus
reprocessed light is relatively enhanced. This leads to a strong
appearance of Fe~K in an observed spectrum (Awaki et al 1991;
Ghisellini, Haardt \& Matt 1994; Krolik, Madau \& Zycki 1994). A large
equivalent width ($EW \sim 1$ keV) of Fe~K is then considered to be a
characteristic feature of heavily obscured AGN with an absorbing
Thomson depth of unity or larger, i.e., Compton-thick AGN.

The goal of this work is to investigate more powerful AGN at higher
redshifts, as detected in the XMM-COSMOS field, to see whether the
above properties hold or evolve with increasing
luminosity/redshift. We show below that the Fe~K properties of AGN in
the XMM-COSMOS field appear different from those in nearby objects,
and discuss possible drivers of these changes.

After presenting an overview of the XMM-COSMOS dataset and the
spectral stacking method employed in this paper, results are presented
for the whole dataset as well as its several subsets in order to investigate
the variations of the Fe~K line properties.



The cosmology adopted here is $H_0=70$ km s$^{-1}$ Mpc$^{-1}$,
$\Omega_{\Lambda}=0.72$, $\Omega_{\rm M}=0.28$.

\section{Observation and data reduction}

About 1800 point-like X-ray sources were detected in the XMM-COSMOS
survey (Cappelluti et al 2009). The mean exposure time over the survey
area is about 40 ks. Their optical identifications and redshifts
compiled from literature, various spectroscopic campaigns (Lilly et
al 2009; Trump et al 2009) and photometric redshifts (Salvato et al
2009, 2011) are presented in Brusa et al (2010). The spectrum extraction
procedure for each XMM-COSMOS source is described in Mainieri et al
(2007), and we use those spectra obtained from the EPIC pn camera.

We use the X-ray data analysis software packages HEASARC's FTOOLS 6.1
and XSPEC version 12 for the analysis presented below.

\section{The Sample}


The X-ray sources used in our spectral stacking analysis are selected
from the XMM-COSMOS catalogue of Brusa et al (2010). 1449
extragalactic sources with a secure optical counterpart (``flagID =
1" in Brusa et al 2010), and either a spectroscopic or photometric
redshift, were selected for further filtering based on the quality of
their X-ray data. Given our interest in the rest-frame 2-10 keV band,
background corrected counts of each source in the rest frame 2-10 keV
band are used to discard poor quality spectra, which would only
increase the noise in a final stacked spectrum. The distribution of
the counts is shown in Fig. 1. A clear break is seen at 20 counts,
which would not appear if observed with infinite sensitivity and thus
this break can be identified as the sensitivity limit (more correctly,
below this limit, the uncertainty in background subtraction becomes too
large, and returns in data stacking diminish). We adopted this break as a
threshold for discarding poor quality spectra of 406 sources. On the
other end, we identified 23 sources with more than 400 counts as
bright outliers (see Fig. 1), and we have separated them from the
general stacking analysis. They are hereafter referred to as the
``Bright 23" sample (see \S~3.3). The exclusion of these brightest
objects leaves 1020 sources with 20-400 source counts. We refer to
these 1020 sources as the main sample, and the results shown below are
derived from this sample unless stated otherwise. The total counts in
the rest-frame 2-10 keV for all sources in the main sample is 77039
cts, half of which come from sources with less than 100 counts
individually.  For sources in the main sample, the typical source
fraction of the accumulated rest-frame 2-10 keV counts is $\sim 50$\%,
i.e., source and background counts are comparable.

\subsection{Main sample} 

In the main sample, 643 sources have spectroscopic redshifts, and
photometric redshifts are available for the other 377 sources. The
median values of redshift are $\tilde{z}_{\rm tot}=1.30$ for all 
sources, $\tilde{z}_{\rm sp} = 1.17$ for the sources with
spectroscopic redshift and $\tilde{z}_{\rm ph}=1.59$ for the sources
with photometric redshift.

For sources with spectroscopic redshifts, those that are broad-line AGN 
based on their optical spectra,  are assigned to ``Class I" , while the 
remainder are assigned to  ``Class II".  Salvato et al (2009, 2011) give the best-fit SED type for individual
sources with photometric redshifts. The photometric sample is divided
into  ``SED I" when the SED class is 19 or larger (those of broad-line AGN,
see Salvato et al 2009 for details), and  ``SED II"  otherwise.

For the spectral stacking discussion below, we divide the sample into two ``Types" 
that are defined as follows: Type I $\equiv $ Class I + SED I, and Type II $\equiv $
Class II + SED II.  In other words, Type I objects are broad-line AGN
or objects with a SED similar to that of broad-line AGN, while Type II
objects are all of the rest, including AGN with  Seyfert~2 excitation,
emission-line galaxies, and absorption-line galaxies. The Type I sample
contains 508 sources with a median redshift of $z=1.60$, while the Type
II sample contains  512 sources with median redshift of $z=0.97$. Their
redshift distributions are shown in Fig. 2. Plots of the 2-10 keV
luminosity (with no correction for absorption) versus redshift for
Type I and Type II objects in the main sample as well as the Bright 23 sample 
are shown in Fig. 3.


\begin{table}
\begin{center}
\caption{Properties of the main and bright samples.}
\begin{tabular}{lccccr}
Class & $N$ & $\tilde z$ & $\tilde L_{2-10}$ & $\alpha_{\rm E} $ & Cts \\
(1) & (2) & (3) & (4) & (5) & (6) \\  
Main & 1020 & 1.30 & 43.91 & $0.60\pm 0.03$ & 77039 \\[5pt]
Spectro-z & 643 & 1.17 & 43.87 & $0.71\pm 0.03$ & 57538 \\
Photo-z & 377 & 1.59 & 43.53 & $0.34\pm 0.05$ & 19501 \\[5pt]
Type I & 508 & 1.60 & 44.16 & $0.80\pm 0.05$ & 48218 \\
Type II & 512 & 0.97 & 43.65 & $0.21\pm 0.05$ & 28821 \\[5pt]
\multicolumn{6}{c}{Type I objects} \\
Class I & 366 & 1.57 & 44.21 & $0.80\pm 0.03$ & 40022 \\
SED I & 142 & 1.78 & 44.03 & $0.59\pm 0.07$ & 8196 \\[5pt]
\multicolumn{6}{c}{Type II objects} \\
Class II & 277 & 0.83 & 43.41 & $0.33\pm 0.07$ & 17516 \\
SED II & 235 & 1.43 & 43.88 & $0.12\pm 0.07$ & 11305 \\[5pt]
\multicolumn{6}{c}{Bright23}\\
Bright23 & 23 & 1.13 & 44.77 & $0.84\pm 0.04$ & 16281 \\
\end{tabular}
\begin{list}{}{}
  Note --- (1) Samples of X-ray sources (see \S~3 for details); (2)
  Number of sources; (3) Median redshift; (4) Median 2-10 keV
  luminosity in logarithmic units of erg s$^{-1}$; (5) Energy index of
  the stacked spectrum; (6) Number of net source counts.
\end{list}
\end{center}
\end{table}

Fig. 4 shows the 2-10 keV luminosity distributions for the Type I and
Type II objects. The median values of logarithmic luminosities are
44.16 and 43.65 (erg s$^{-1}$), respectively.


\begin{figure}
\centerline{\includegraphics[width=0.4\textwidth,angle=0]{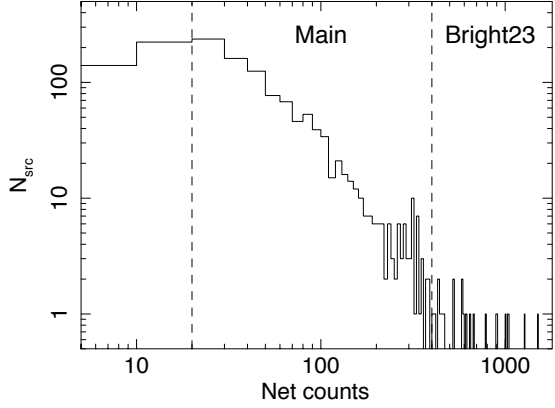}}
\caption{Distribution of the background-corrected counts of individual
  spectra in the rest-frame 2.1-10.1 keV band, obtained from
  XMM-Newton EPIC pn.}
\end{figure}


\begin{figure}
\centerline{\includegraphics[width=0.4\textwidth,angle=0]{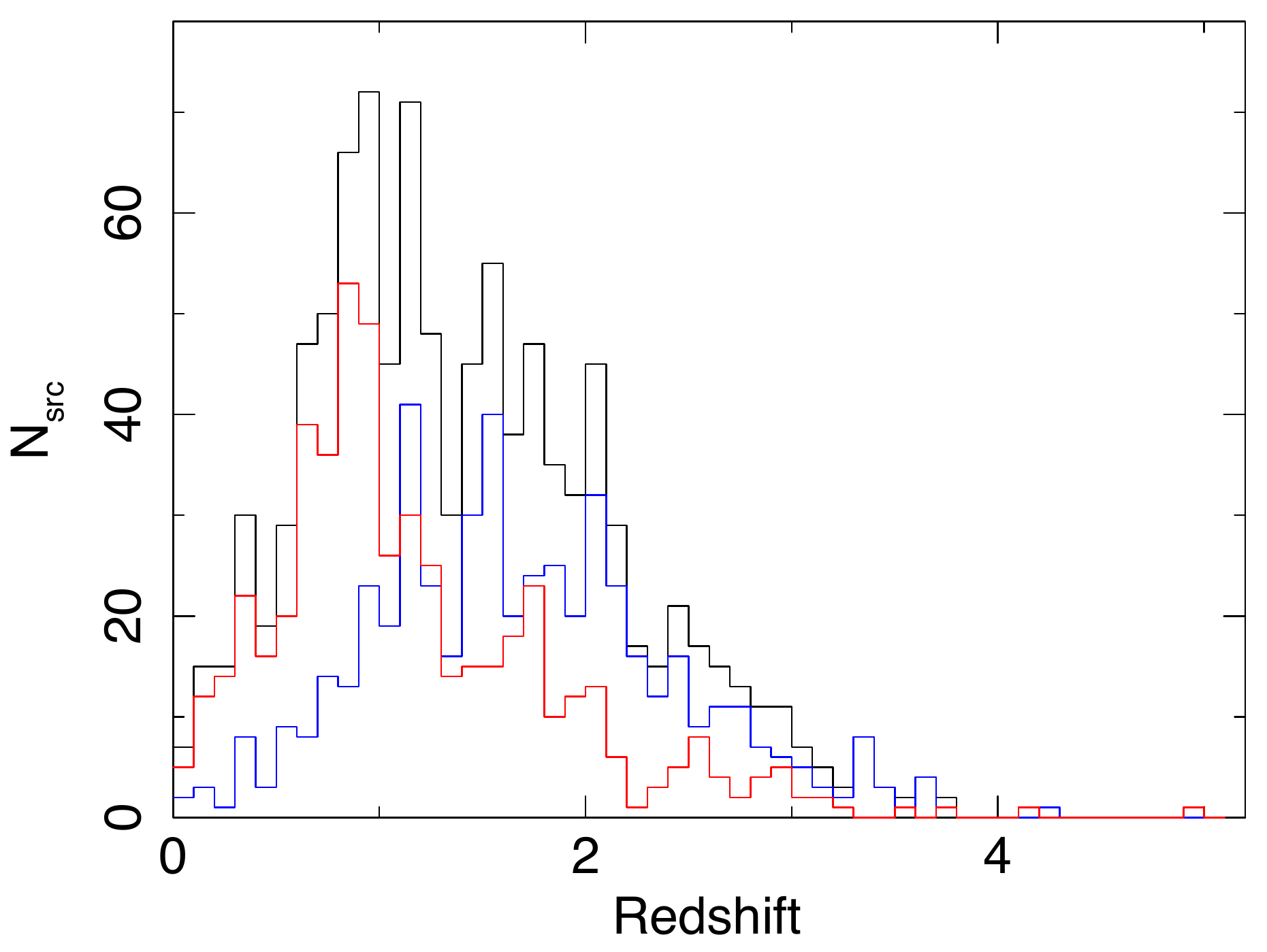}}
\caption{Distribution of redshifts for 1020 sources in the main sample
  (black). The Type I objects (blue) and Type II objects (red) are
  also shown separately. The median redshifts for the total, Type I
  objects and Type II objects are 1.30, 1.60, 0.97, respectively.}
\end{figure}


\begin{figure}
\centerline{\includegraphics[width=0.4\textwidth,angle=0]{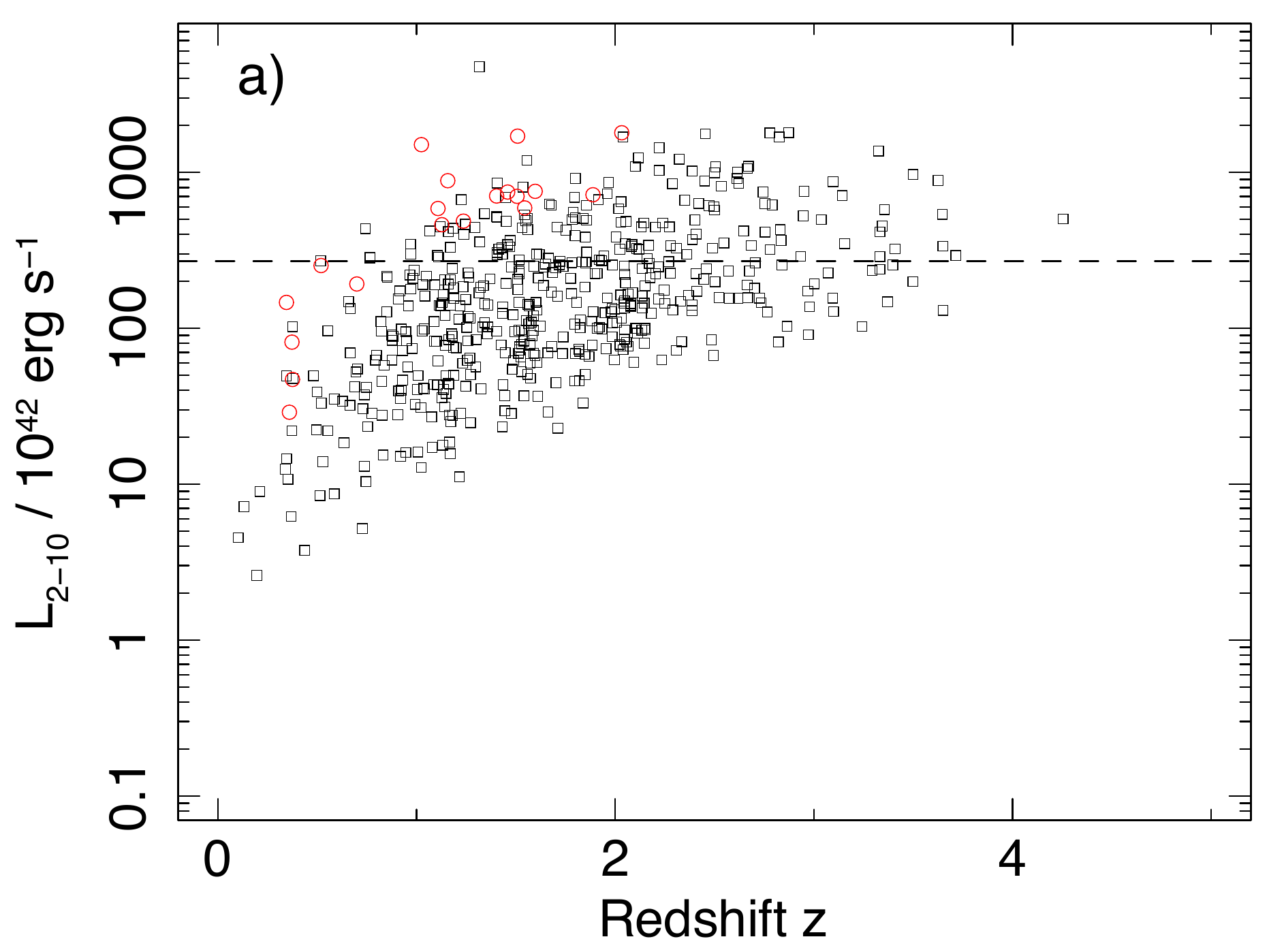}}
\centerline{\includegraphics[width=0.4\textwidth,angle=0]{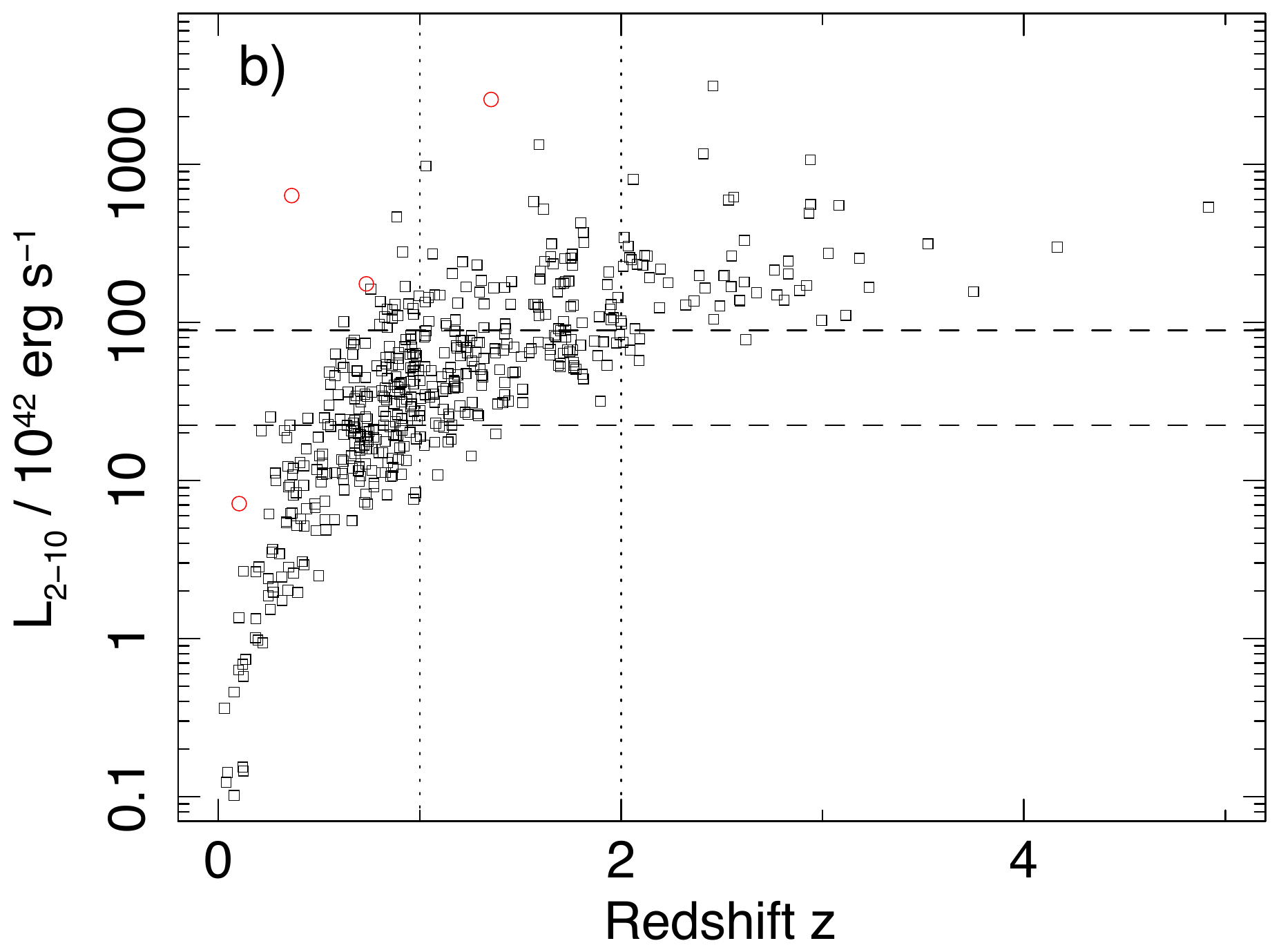}}
\caption{Distribution of  sources in the rest-frame 2-10 keV
  luminosity--redshift plane for a) Type I and b) Type II
  objects. Sources in the main sample are indicated in black and the
  23 sources in the bright sample are in red. Luminosity boundaries
  for investigating spectral variations are indicated by dashed lines
  (\S~5.1.1, \S~5.2.1). Redshift boundaries at $z=1$ and $z=2$
  are also indicated for the Type II objects.}
\end{figure}


\begin{figure}
\centerline{\includegraphics[width=0.4\textwidth,angle=0]{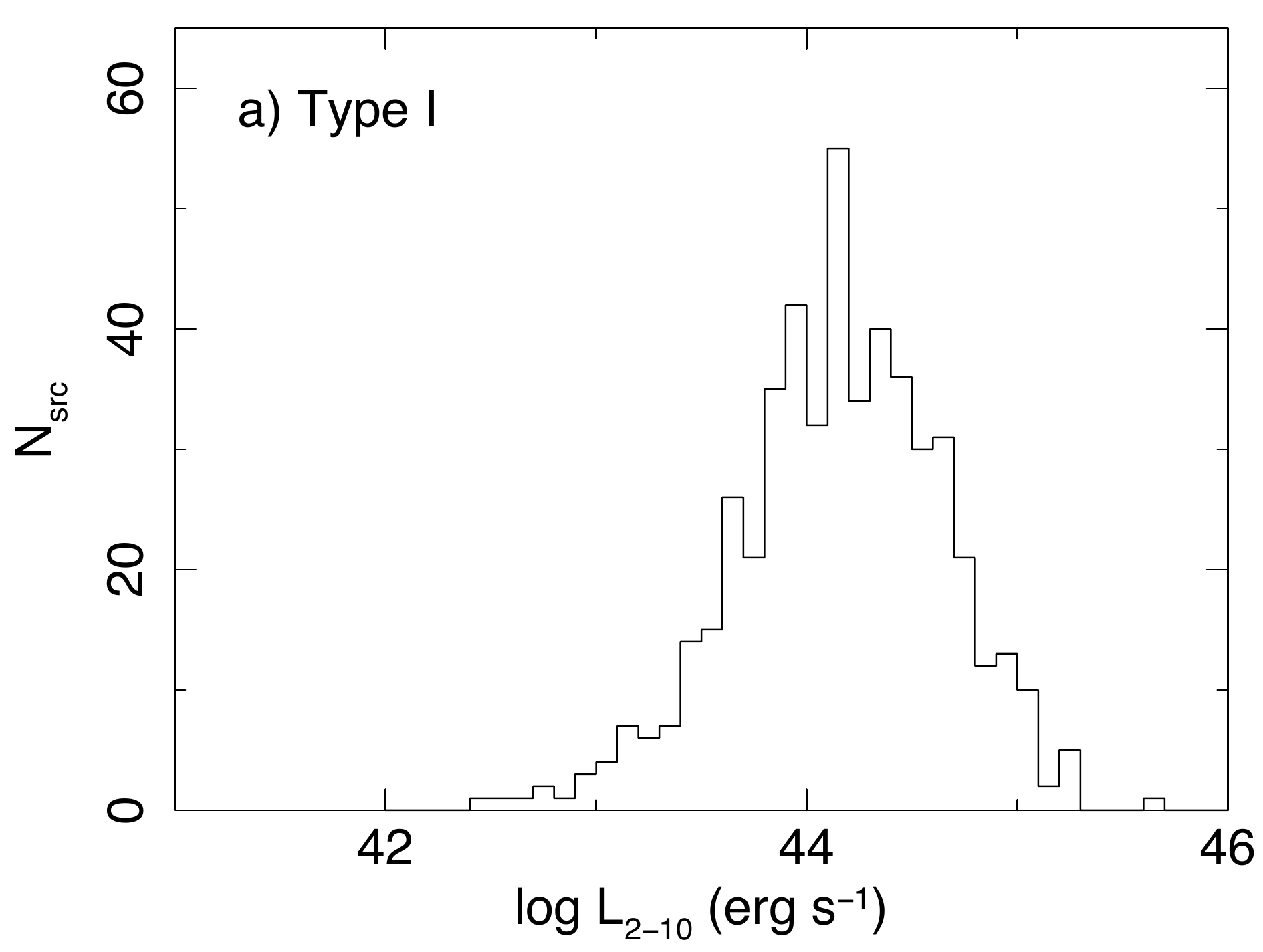}}
\centerline{\includegraphics[width=0.4\textwidth,angle=0]{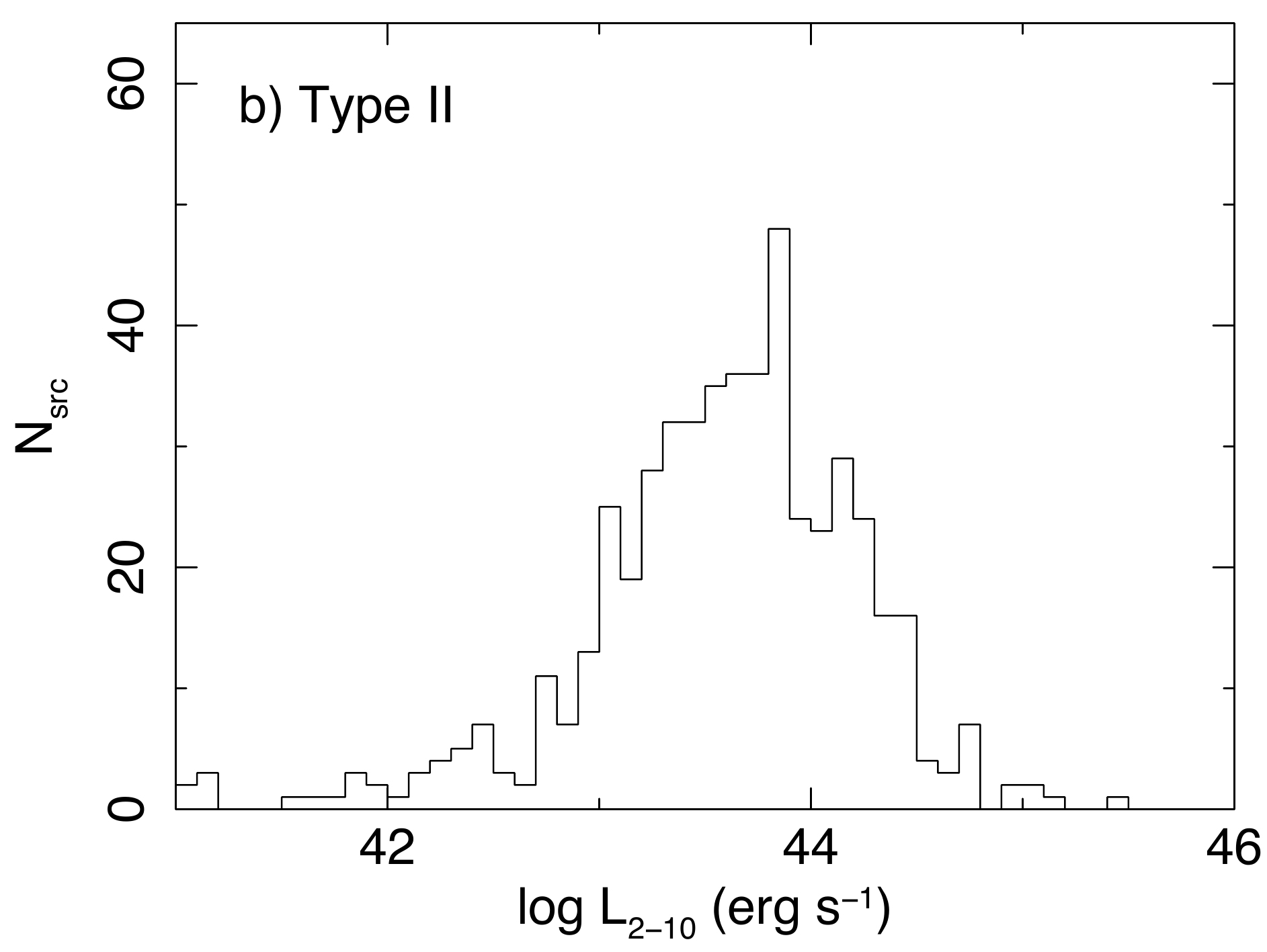}}
\caption{Distribution of the rest-frame 2-10 keV
  luminosity of the sample sources: a) Type I,  and b) Type II
  objects.}
\end{figure}

\subsection{Broad-line AGN: $M_{\rm BH}$ and $\lambda $}

For spectroscopically confirmed broad-line AGN (Class I), their black
hole masses estimated from the optical broad lines are available from
the literature (Trump et al 2009; Merloni et al 2010). Merloni et al
(2010) use Mg {\sc ii} while Trump et al (2009) give estimates based
either on C {\sc iv}, Mg {\sc ii} or H$\beta$ where available. There
are 37 objects for which both papers have measurements based on Mg
{\sc ii}. A comparison between the two papers finds no systematic difference
in $M_{\rm BH}$ values, but there is a scatter of $\sim 0.3$ dex, which can be
considered as the uncertainty in $M_{\rm BH}$ estimates. In the analysis of
Trump et al (2009), $M_{\rm BH}$ based on Mg {\sc ii} is considered to be most robust, and  
in the event that a  Mg {\sc ii} based value is not available, estimates of $M_{\rm BH}$ 
based on other emission lines are adopted.  Out of 366 Class I AGN in the
main sample, 181 objects are found to have 
$M_{\rm BH}$ estimates.  Given the  availability of  spectral lines, a
large fraction of these objects (149) lie in the redshift range
$z=$1-2.2.

The accretion rate relative to the classical Eddington limit (against
electron scattering), i.e. the Eddington ratio, $\lambda $, is
estimated from a ratio of the bolometric luminosity ($L_{\rm bol}$)
and the Eddington luminosity ($L_{\rm Edd} = 1.3\times 10^{38}(M_{\rm
  BH}/M_{\odot})$ erg s$^{-1}$). For almost all Class
I AGN with determined $M_{\rm BH}$ (177 out of 181), their bolometric luminosities were computed 
by Lusso et al (2010) using the multi-wavelength SED. For these 177
objects, the range of log $(M_{\rm BH}/M_{\odot})$ is between 7.18 and
9.70 with a median value of 8.39 (Fig. 5a) while  log ($\lambda$) 
lies in the range $-1.9$ to $+0.4$ with a median value of $-1.0$
(Fig. 5b). There is a weak inverse correlation between the
accretion rate and black hole mass (see also Trump et al 2009; Netzer
\& Trakhtenbrot 2007), but with a large scatter, which is partly due to
a selection bias in a flux-limited sample. This subset has a median
redshift, $\tilde z=1.51$, and median 2-10 keV luminosity log~$(L_{2-10}) = 44.26$ \ergps, 
and median bolometric luminosity log~($L_{\rm bol}) = 45.52$ \ergps.


\begin{figure}
\centerline{\hbox{
\includegraphics[width=0.25\textwidth,angle=0]{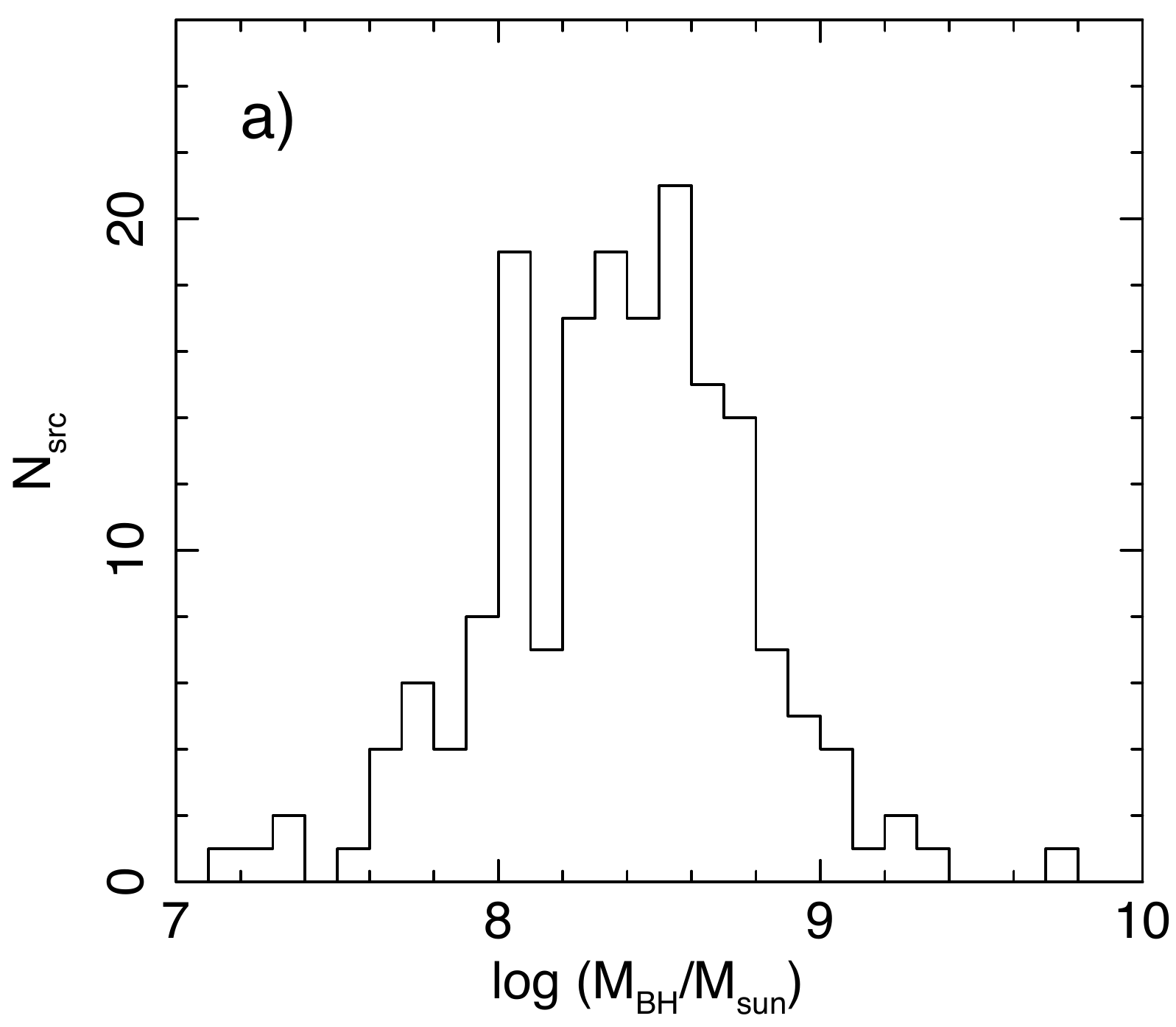}
\includegraphics[width=0.248\textwidth,angle=0]{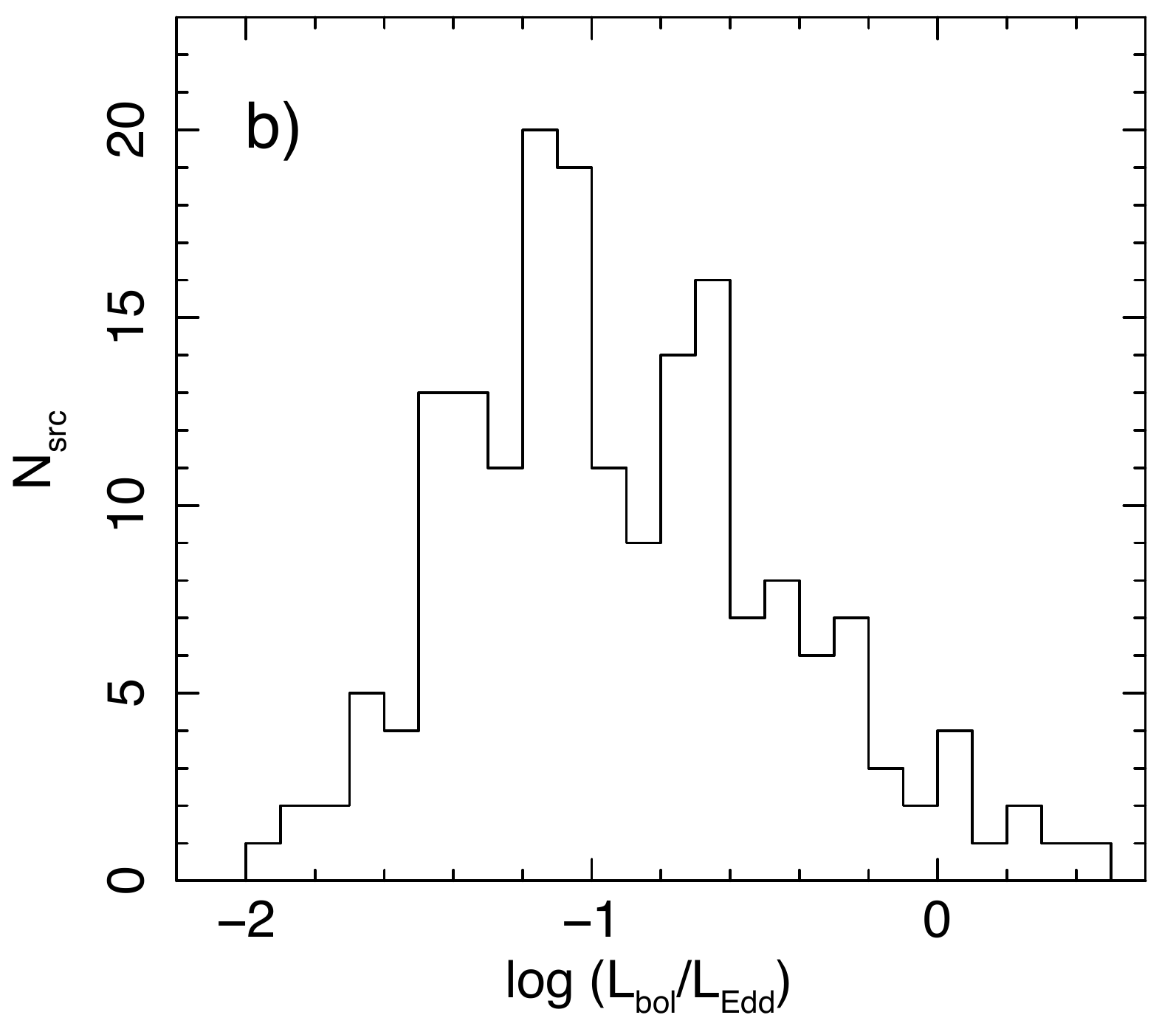}}}
\caption{Left: The distribution of the black hole masses estimated
  for the spectroscopically confirmed Type I AGN (177 objects) taken
  from Trump et al (2009) and Merloni et al (2010). The median value
  is 8.39. Right: The distribution of the Eddington fraction $\lambda $
  for those Type I AGN. The median value is $-1.0$. }
\end{figure}

\subsection{Bright 23}

The 23 sources with more than 400 net counts each in the rest frame 2-10 keV 
band are mostly broad-line AGN (19 Type I objects, 4 Type II objects) at $z\leq 2$.  
In Fig. 3, they are found in the upper envelope of the source distribution and 
represent the most luminous objects at a given redshift. The median properties are given in Table 1.


\section{Stacking method}


\begin{figure}
\centerline{\includegraphics[width=0.47\textwidth,angle=0]{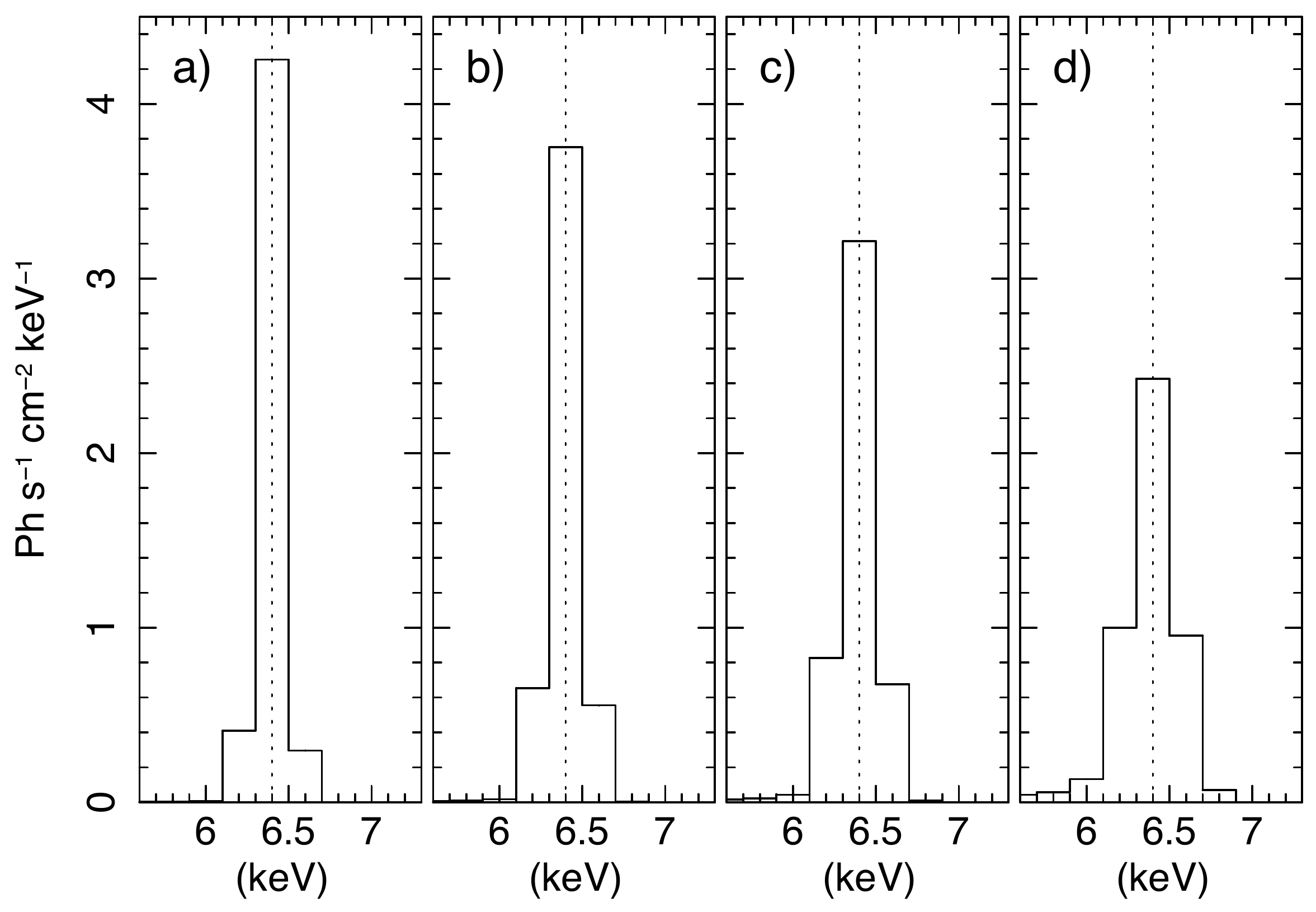}}
\caption{A narrow Gaussian was simulated at redshifts;  a) $z = 0$, b)
  0.5, c) 1 and d) 2, as observed with the EPIC pn, and then converted
  to a rest-frame spectrum after  correcting for  detector
  response and  redshift  for  individual sources in
  the stacking sample. The rest-frame line energy is accurately
  recovered. The line profiles of higher redshift objects broaden
  because the energy resolution, ($\Delta E/E$), of the detector becomes
  worse at lower energies. For this reason, while the integrated line
  flux is recovered, the line peak is lowered.}
\end{figure}

The rest-frame 2.1-10.1 keV data for individual sources are
stacked. An original EPIC pn spectrum has 4096 channels with 5 eV
channel width. The rest-frame spectral stacking is designed to have a
2.1-10.1 keV rest-frame energy range with 200 eV channel width. In an
individual source with a redshift $z$, the EPIC pn data in the
[2.1/(1+z)] keV to [10.1/(1+z)] keV range are selected and divided
into 40 energy intervals with [200/(1+z)] eV. These data, after
background subtraction, are corrected for the detector response curve
and for Galactic absorption, \nH = $2.6\times 10^{20}$ \psqcm
(Kalberla et al 2006), before correcting the energy scale for
redshift.

The data stacking is a straight sum of individual sources. The reason
why normalizing each spectrum by its flux was not made was the very large
uncertainty in choosing the proper normalizing factor for faint sources, e.g., those with 
$\sim 20$ counts, which represent the majority of the sample. Thus the stacked spectra
presented below naturally give more weight to brighter
sources.  However, our primary objective is to detect faint signals of
Fe~K emission in distant active galaxies that cannot reliably be detected by
an ordinary observation of a single source, rather than to obtain a ``mean" 
spectrum.

As the energy resolution, $\Delta E/E$, of the EPIC pn degrades
towards lower energies (approximately $\propto E^{-0.5}$), a
redshift-corrected line profile of a higher redshift object would be
broader than that of a lower redshift object. This effect is demonstrated in
Fig. 6. Narrow Gaussians (with dispersion $\sigma = 1$ eV) were
simulated using an identical normalization, but at different redshifts, $z=
$0, 0.5, 1, and 2, and then the same correction procedure described
above was applied to plot them in rest-frame energy.

The energy resolution of the EPIC pn is $\approx 0.15$ keV (FWHM) at
6.4 keV. However, given the design resolution of 0.2 keV, a simulated
narrow line profile for z=0 -- (a) has FWHM = 0.21 keV corresponding to a  Gaussian
dispersion, $\sigma = 0.09$ keV. As  redshift increases, the line
profile broadens to $\sigma = 0.15$ keV at  $z=2$. While
the Fe~K line intensity distribution as a function of redshift is, of
course, not known, if we assume that  the median value of redshift is 
representative, the artificial line broadening due to  instrumental
effects would be $\sigma \approx 0.094$ keV for the stacked spectrum of
Type I AGN, and $\sigma \approx 0.076 $ keV for Type II
AGN.   Thus, our integrated spectra are insensitive to a line width with
$\sigma \leq 0.1$ keV in the Fe~K band. The Fe lines discussed below
are unresolved in most cases, otherwise the given resolution has been 
corrected for  artificial broadening, assuming a typical redshift.

\section{Results}

For convenience of visual inspection, stacked spectra analyzed in this
paper are presented in units of flux density ($F_{\rm E}$), and the 
spectral slope refers to an energy index, $\alpha_{\rm E}$,  defined
as $F_{\rm E}\propto E^{-\alpha_{\rm E}}$.   The photon index, $\Gamma $, commonly
used in X-ray astronomy, is related to the energy index as $\Gamma =
1+\alpha_{\rm E}$.   Fig. 7 shows integrated (stacked) spectra for 4 subsets of objects:\ 
(1) the  main sample, (2) the ``Bright~23",  (3) Type I objects, and  (4) Type II objects.

The spectrum of the main sample has a spectral slope, $\alpha_{\rm E}
= 0.60\pm 0.03$ (hereafter errors quoted to spectral parameters are of
$1\sigma $). This continuum represents a sum of all the sources in the
rest-frame 2-10 keV band, and is steeper than the value of
$\alpha_{\rm E}\simeq 0.4$ for the X-ray background in the observed
2-10 keV band. A likely explanation for this difference is that
bright, less obscured sources dominate the total integrated spectrum,
given the sensitivity limit of the XMM-COSMOS. A significant Fe~K
feature is found at $\sim 6.5$~keV.  Fitting the Fe~K excess with a
single Gaussian gives a line centroid of $6.55\pm 0.16$ keV with
$\sigma = 0.3^{+0.2}_{-0.1}$ keV and an equivalent width, $EW =
0.18\pm 0.07$ keV. When modelled with two Gaussians, which is more
realistic, the line centroids are $6.36^{+0.09}_{-0.06}$ keV and
$6.86^{+0.11}_{-0.12}$ keV, with a flux ratio of 1:0.8. The former is
unresolved and can be identified with cold Fe~K. The latter is
marginally resolved with $\sigma = 0.15^{+0.2}_{-0.1}$ keV, and
together with the centroid energy, this suggests a blend of Fe {\sc
  xxv} at 6.70 keV and Fe {\sc xxvi} at 6.97 keV. No clear sign of a
broad red wing is found.

The integrated spectrum of  the ``Bright23" sample shows a power-law spectrum with
$\alpha_{\rm E} = 0.84\pm 0.04$. No significant Fe~K emission is
detected. If anything, a broad emission feature with $EW \sim 0.08$ keV might be
present, but its excess above the continuum is barely detected at the $1\sigma$
level. The $2\sigma $ upper limit for a narrow line at 6.4 keV
corresponds to $EW=0.09$ keV.

The spectra of Type I and Type II objects are discussed further in the
following subsections. The spectrum of Type II objects is
significantly harder ($\alpha_{\rm E}\simeq 0.2$), as expected for
various absorbed continua that are superposed, rather than that of Type I
objects ($\alpha_{\rm E}\simeq 0.8$).

To assess the effect errors in the photometric redshifts, the stacked
spectrum for SED II sources is inspected, since these objects are
expected to emit a narrow Fe~K line. According to Salvato et al
(2009), the accuracy of the photometric redshifts is estimated to be
$\simeq 1.5$ \%, which can be translated to $\sigma\simeq 0.1$ keV for
a narrow emission-line at 6.4 keV. The stacked spectrum for SED
II sources shows an unresolved ($\sigma = 0.1^{+0.4}_{-0.1}$ keV) line
at $6.38\pm 0.13$ keV with $EW =0.10\pm 0.07$ keV. Relative to the
spectroscopic sample of Type II objects (Class II), the EW of the 6.4
keV emission is $\sim 60$\%. Given the loose constraint on the line
width, large errors in photometric redshifts cannot be ruled out as
the cause of the reduction in the EW of Fe~K.  However, the best-fit
line width agrees with the expected line broadening from the error
reported in Salvato et al (2009), and the redshift/luminosity
dependence (see \S~5.2) may be a more likely cause, as the SED II
sample includes more luminous, higher redshift objects than the Class
II sample (see Table 1).


\begin{figure}
\centerline{\includegraphics[width=0.5\textwidth,angle=0]{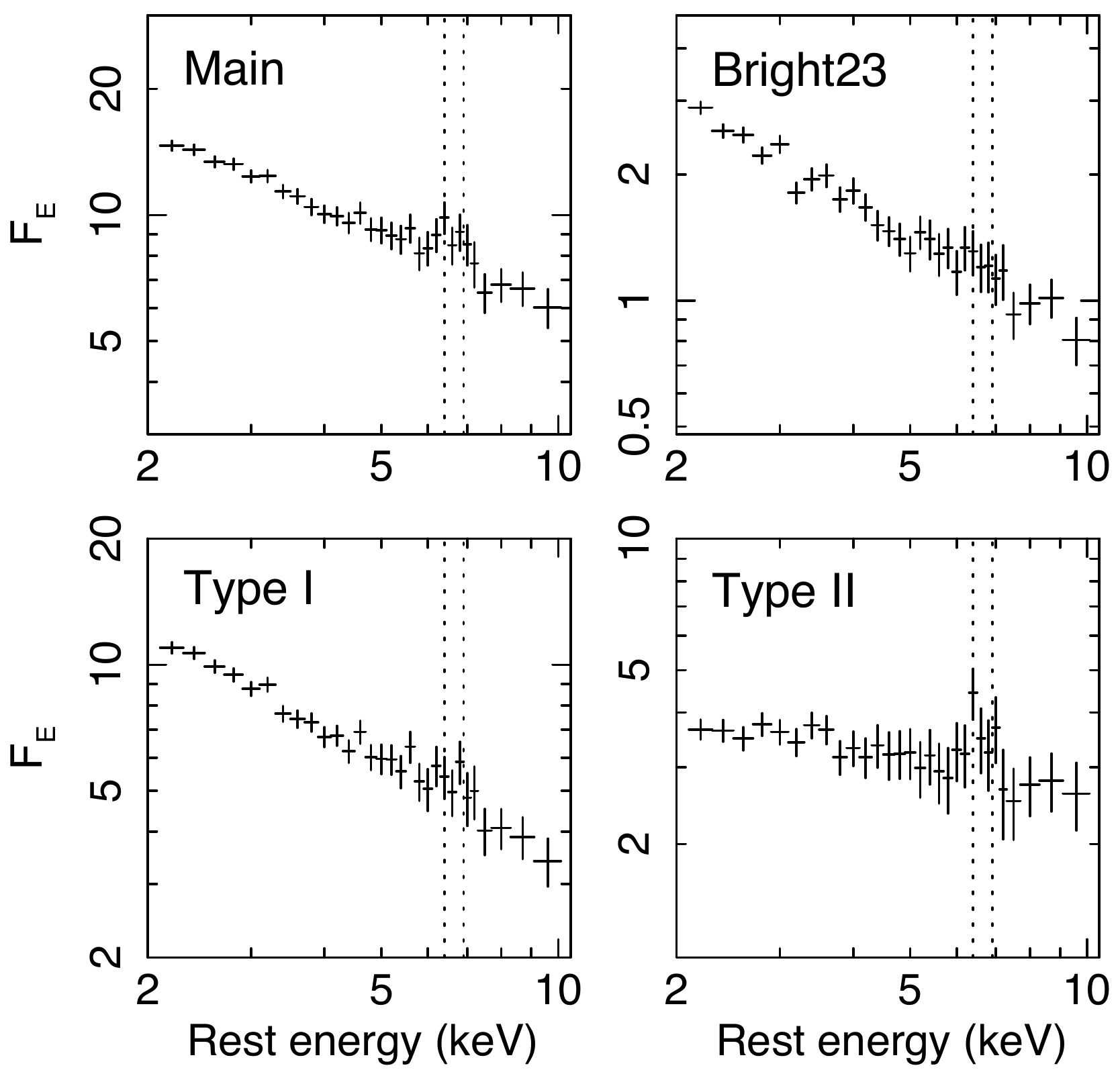}}
\caption{Stacked spectra of the main sample, ``Bright23" sample, Type I
  objects and Type II objects. The spectra are presented in flux
  density  units of $10^{-13}$ [erg cm$^{-2}$ s$^{-1}$ keV$^{-1}$]. The rest energies of 6.4 keV and 6.9 keV are indicated by dotted lines.}
\end{figure}



\subsection{Type I AGN}

Type I objects consist of 366 sources with spectroscopic redshifts
(Class I) and 142 sources with photometric redshifts (SED I). The
sources in the photometric sample are, on average, fainter than those
in the spectroscopic sample, and their contribution to the stacked
spectrum is 20\%. 

The 2-10 keV continuum has a slope,  $\alpha_{\rm E} = 0.80\pm 0.05$,
similar to the average value observed in nearby Seyfert 1 galaxies and
QSOs observed with XMM-Newton, e.g., the CAIXA sample (Bianchi et al
2009). The spectrum of the photometric sample shows a harder continuum, 
$\alpha_{\rm E} = 0.59\pm 0.07$,  than that of the spectroscopic sample, 
$\alpha_{\rm E} = 0.80\pm 0.03$. This suggests that the SED I sample
contains some absorbed sources that might be mis-classified as Type I
AGN. The contamination rate was estimated to be $\sim 12$ \% in Lusso
et al (2010) for the objects for which both spectroscopic and
photometric redshifts are available.

There is excess emission at $\sim 6.6$~keV (Fig. 7), which appears to
consist of two unresolved emission-line components at $6.4\pm 0.2$ keV
and $6.8\pm 0.2$ keV rather than a single broad line. The line flux
ratio of these two lines is approximately 2:3. The EW of the lines are
$0.04\pm 0.02$ keV and $0.05\pm 0.02$ keV, respectively. With a median
2-10 keV luminosity of log $L_{2-10}\simeq 44.2$ erg s$^{-1}$, the EW
of the 6.4 keV line agrees with that expected from the
Iwasawa-Taniguchi (IT) relation (Iwasawa \& Taniguchi 1993) found for
the CAIXA sample (Bianchi et al 2007). The 6.8 keV component is likely
a blend of Fe {\sc xxv} (6.70 keV) and Fe {\sc xxvi} (6.97 keV). These
high-ionization lines are often observed in nearby AGN (e.g., Bianchi
et al 2009) but they are usually weaker than the 6.4 keV line. This is
reversed in our stacked spectrum of more powerful AGN at higher
redshift.　The nature of the enhanced high-ionization line is examined
in \S~5.1.2.

No clear sign of redward broadening in the Fe~K line profile, as
observed in nearby Seyfert galaxies (e.g., Tanaka et al 1995), has
been found either in the stacked data of the Type I objects,
the Class I objects, or the SED I objects. However, an upper limit for a
relativistically broadened Fe~K line with a shape typically found in
nearby AGN (e.g., de la Calle P\'erez et al 2010), is $\sim 0.2$ keV.

With regard to Fe~K emission in Type I objects, we would like to
highlight the following two points, which warrant further discussion 
since they may illustrate possible differences in spectral
characteristics between the distant population of active galaxies, as
sampled by XMM-COSMOS, and nearby active galaxies: 1) No strong
line-broadening, such as might  result from relativistic effects is
found; and 2) The high-ionization lines  Fe {\sc xxv} and Fe {\sc
  xxvi} increase in strength.

In order to search for a population that emits strong broad iron
lines, or the high-ionization lines, we explore the Type I dataset by
dividing it into three subsets according to the 2-10 keV
luminosity, redshift, and black hole mass or Eddington ratio. Here we
report our findings in terms of the above two points.

\subsubsection{Broad Fe~K line}

\begin{figure}
  \centerline{\includegraphics[width=0.35\textwidth,angle=0]{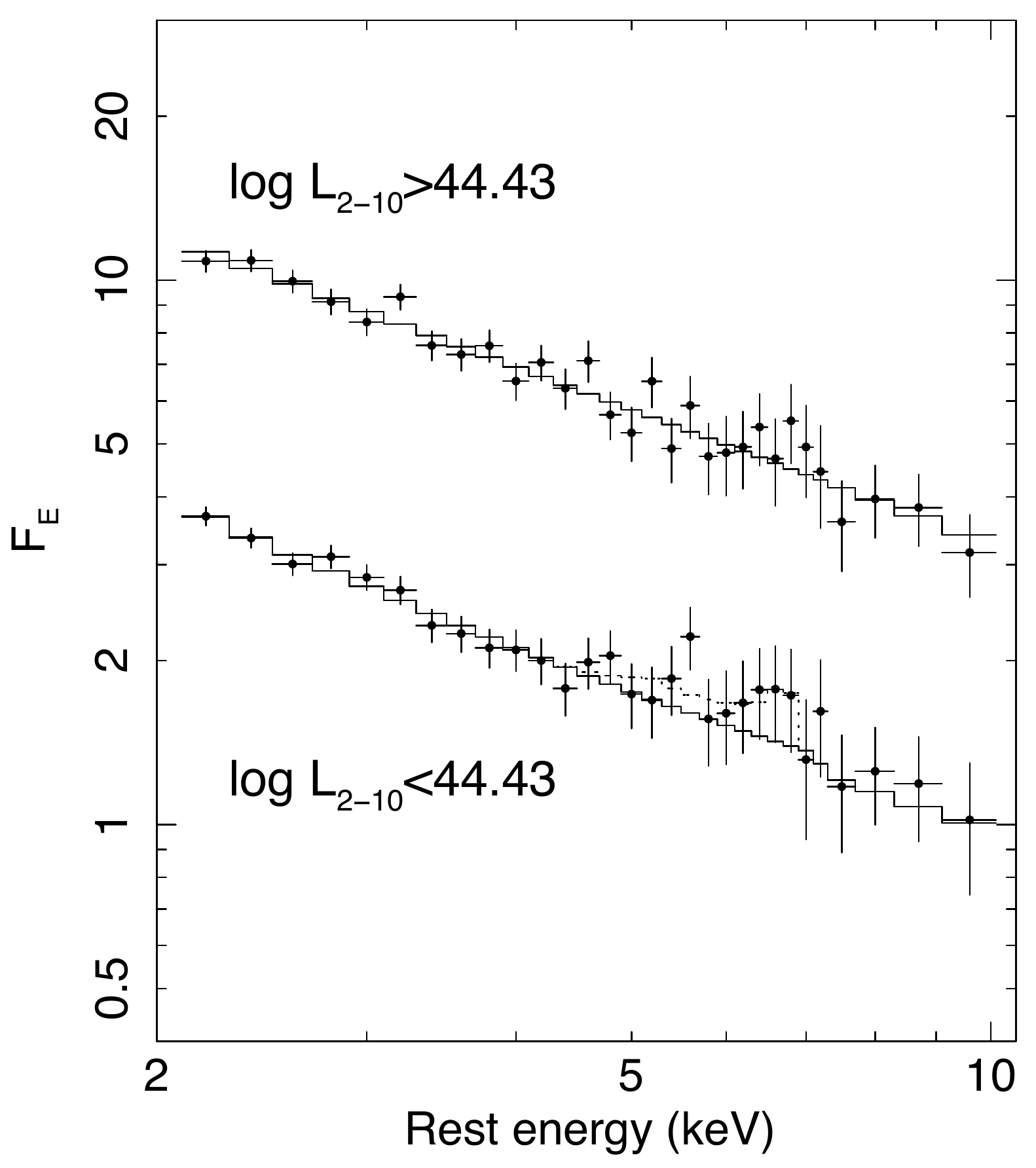}}
  \caption{Stacked spectra of Class I objects (BLAGN) in the
    luminosity ranges above and below log $L_{2-10} = 44.43$ (erg
    s$^{-1}$). For the spectrum of the high luminosity objects, the
    best-fit power-law continuum is indicated by the solid-line
    histogram. Weak Fe~K emission peaking at 6.86 keV is seen. On the
    other hand, the continuum model indicated with the solid-line
    histogram for the spectrum of low luminosity objects is that expected for a 
    power-law with reflection (see text for detail). A moderately
    broadened Fe~K feature centred at 6.5 keV, with a possible redward
    extension down to 4.5 keV, can be seen. The best-fitting diskline
    model for this Fe~K feature is indicated by the dotted-line
    histogram.}
\end{figure}

To avoid a possible artifact due to errors in redshift estimate, which
could result in a false detection of line broadening, we discarded
objects with photometric redshifts and used only the spectroscopic
sample (Class I) for the further investigation of broad emission.
Empirically, it has been found that at least $\sim$150,000 counts in
the 2-10 keV band are required for a significant detection of a
broad Fe~K line with $EW\sim 0.1$-0.15 keV, typical of that found in nearby
Seyfert galaxies, with the EPIC camera (de la Calle P\'erez et al 2010). The
integrated counts of the entire sample of Class I objects is $\sim$40,000,
well below the critical counts (the background is also much higher), and only strong broad emission, with
$EW \geq $0.2-0.3 keV, can be detected. Bearing this limitation in mind,
nevertheless, it is found that there may be a favourable range in
X-ray luminosity and/or in Eddington ratio for finding broad Fe~K
emission, the details of which are described below. 

When the Class I objects are divided into two X-ray luminosity ranges,
above and below $3\times 10^{44}$ erg s$^{-1}$, a possible signal of a
broadened Fe~K emission line is found in the lower X-ray luminosity
subset, albeit the significance of the broad component is barely at
$2\sigma $ level. This subset consists of 247 objects with median
redshift $\tilde z = 1.45$ and median 2-10 keV luminosity, log
($L_{2-10}) = 44.00$ erg s$^{-1}$.

The stacked spectrum is shown in Fig. 8. The continuum slope is
$\alpha_{\rm E}=0.88\pm 0.05$. Fitting a single Gaussian to the Fe~K
feature gives a line centroid of $6.55\pm 0.26$ keV with $\sigma =
0.37^{+0.58}_{-0.16}$ keV. This simple fit leaves a weak but positive
residual down to 4.5 keV, suggesting the presence of a red wing. For
comparison, the spectrum of objects in the higher luminosity range is
also shown. The spectrum of the high luminosity subset agrees well
with a simple power-law of $\alpha_{\rm E} = 0.80\pm 0.05$ apart from narrow
emission-line features at $6.86\pm 0.15$ keV and possibly at $6.38\pm
0.21$. There is no hint of excess emission below 6 keV. This subset
carries a factor of 1.4 more flux than the lower luminosity subset with the broad feature, which would explain why no broad emission was
seen in the total spectrum.

While the line broadening is marginally significant, presence of a red
wing below 6 keV is highly uncertain and introducing any further
complexity in a spectral fit will be subject to
overmodelling. However, in the interest of a comparison with the
nearby AGN, we present a modelling of the broad feature, assuming that
it is an emission line arising from a relativistic disk. A reflection
continuum could mimic a weak red-wing and its effect appears at
maximum without relativistic smearing, as modelled with {\tt pexrav}
(Magdziarz \& Zdziarski 1995). The broad feature is found to be robust
aginst this. In the continuum model, the cold reflection is included
with a reflection fraction of unity and inclination of the reflecting
surface to be $35^{\circ }$ (see below). The line feature is modelled
by the relativistic line model {\tt diskline} (Fabian et al
1989). Given the quality of our data and the degenerate nature of some
model parameters, typical values for the disk-line model were chosen
and only the disk inlination and the line normalization are
fitted. The disk radii were assumed to be between 10-20 $r_{\rm g}$
(see Nandra et al 2007 for the typical radius) with an emissivity
index of $-2$, and the line was assumed to be emitted at 6.4
keV. Combined with the continuum model including reflection, the
best-fit inclination is found to be $35^{\circ}\pm 7^{\circ}$. The
line flux corresponds to $EW = 0.28^{+0.20}_{-0.14}$ keV. This model
reproduces the moderately broad 6.55 keV peak, picked up by the single
Gaussian fit and the possible redward extension down to 4.5 keV. No
rapid spin of the black holes is inferred from the redwards extension
of the line in the present data, but much higher quality of data in
the red-wing band are needed to impose a meaningful limit on the spin.

The above EW is likely an upper limit for any  broad component of
the line, since the emission peak at 6.55 keV is at least partly due to a
blend of the cold Fe line at 6.4 keV and high-ionization lines at
6.7-7 keV. It has been known that a narrow Fe~K line at 6.4 keV,
originating from cold matter at large distances, is ubiquitously found
in AGN. Using the median X-ray luminosity of the lower luminosity subset, combined with
the known IT relation (Bianchi et al 2007), the EW of the 6.4 keV line is
expected to be $\sim 0.05$ keV. Furthermore, a contribution by the
high-ionization lines with $EW \sim 0.05$ keV is also expected at
energies corresponding to  the blue wing. Thus, if present, the true contribution of the broad
component would be $EW\leq 0.2$ keV. Note also
that the subtraction of the non-relativistic components affects the
diskline parameter, in particular the inclination, which would be
lowered.

A similar broad emission feature can also be found when an
intermediate range of the Eddington ratio, e.g. $\lambda\sim 0.1$, is
selected (Fig. 9). The details of the source selection according to
$\lambda $ is presented in Table 2. Although the signal to noise ratio
is lower, given that the number of objects in the stack is only 97,
besides a narrow line at $6.50\pm 0.02$ keV, a red wing can be seen
down to 5 keV. The level of detection of this broad feature is also
$\simeq 2\sigma $. The typical X-ray luminosity (log $L_{2-10} =
44.17$ \ergps ) of this subset is within the low X-ray luminosity
range (log $L_{2-10}\leq 44.43$ \ergps) where the broad emission seems
to be present. This makes it difficult to determine which parameter
drives the production of possible broad Fe~K emission. 

In summary, a broad red wing of Fe~K emission in Type I AGN may be
present, when the X-ray luminosity range is restricted to below
$\simeq 3\times 10^{44}$ \ergps, or when sources with an intermediate
range of Eddington ratio ($-1.3< {\rm log}\thinspace (\lambda ) \leq
-0.6$) are selected. The detection of the broad emission is, however
of low significance, and the EW of the broad feature would be $\leq
0.2$ keV, in agreement with the previous studies by Corral et al
(2008) and Chaudhary et al (2011). Note that our objects lie, on
average, at higher redshifts than those in the previous work, and the
sample of Chaudhary et al (2011) contains both Type I and Type II AGN,
which makes a direct comparison difficult but an overall agreement on
broad emission is found. As argued above, our data are expected to be
only sensitive to strong broad-lines with $EW\geq 0.2$-0.3 keV, and
the obtained limit is as expected.


\begin{figure}
\centerline{\includegraphics[width=0.35\textwidth,angle=0]{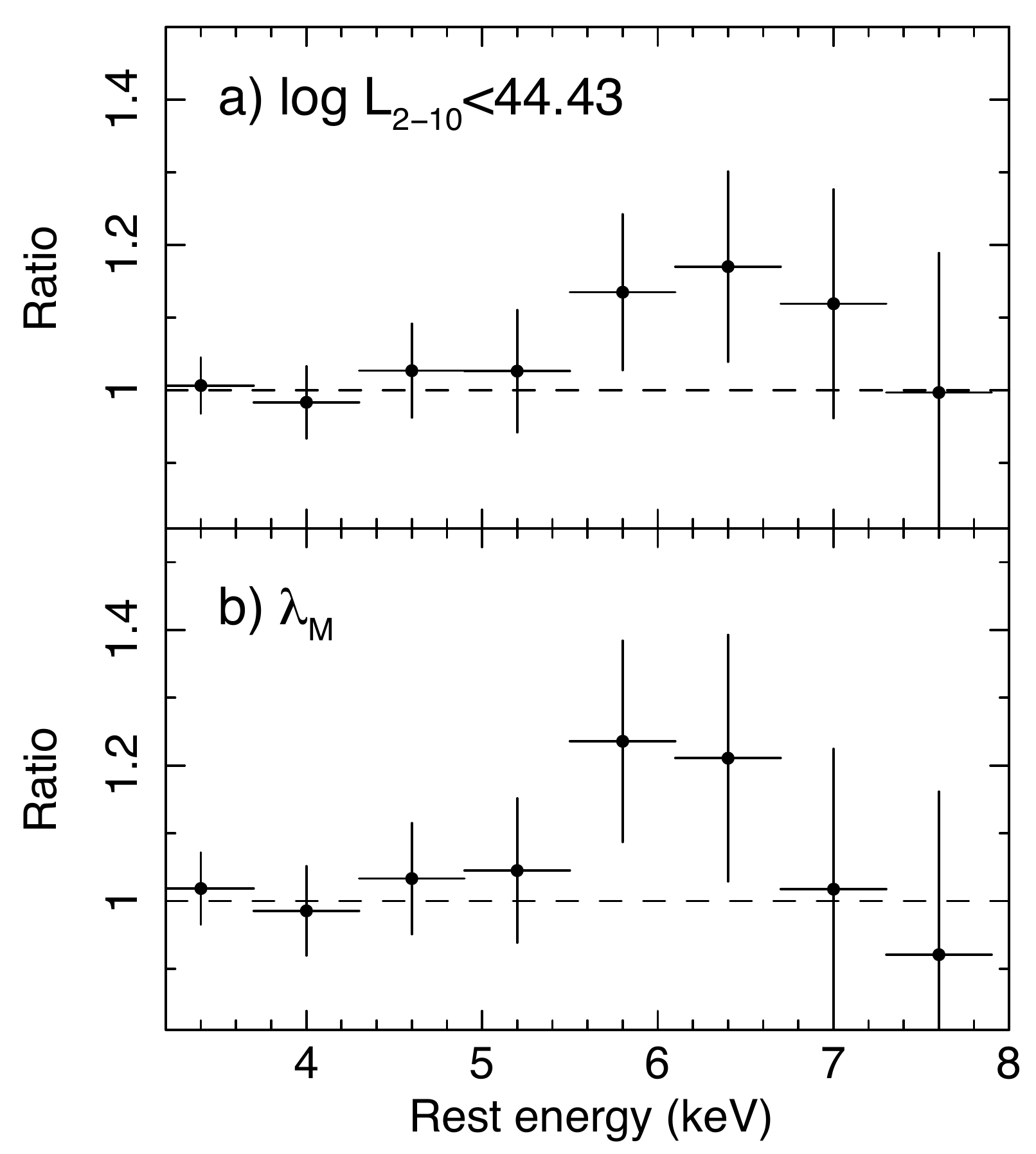}}
\caption{Possible broadened Fe line profiles found in spectra of a)
  lower luminosity range in Class I objects and of b) the intermediate
  range of Eddington ratio (Table 2 and Fig. 10b), presented in the
  form of ratio versus the power-law continuum. The data are rebinned
  to 600 eV intervals for display purposes. }
\end{figure}

\subsubsection{High-ionization Fe emission}

\begin{table*}
\begin{center}
  \caption{Properties of Class I objects in three ranges of accretion rate, $\lambda$.} 
\begin{tabular}{lcccccccccc}
Subset & Range & $N$ & $M_{\rm BH}$ & $\lambda$ & $z$ & $L_{2-10}$ & $F$ & $F/N$ & $\alpha_{\rm E}$ & Fe~K \\
(1) & (2) & (3) & (4)  & (5)  & (6) & (7) & (8) & (9) & (10) & (11) \\
$\lambda_{\rm L}$ & $\leq -1.3$ & 40 & 8.51 & $-1.48$ & 1.20 & 44.18 & 4.2 & 1.1 & $0.87\pm 0.10$ & -- \\  
$\lambda_{\rm M}$ & $-1.3$ - $-0.6$ & 97 & 8.39 & $-1.00$ & 1.51 & 44.17 & 9.3 & 0.96 & $0.86\pm 0.07$ & Broad? \\ 
$\lambda_{\rm H}$ & $>-0.6$ & 39 & 8.10 & $-0.31$ & 1.82 & 44.52 & 4.5 & 1.2 & $0.77\pm 0.07^{a}$ & 6.9 keV \\
\end{tabular}
\begin{list}{}{}
\item[Note]---177 objects with lambda estimates are divided into three
  subsets according to their value of $\lambda $. One source XID 5230 in the
  $\lambda_{\rm H}$ subset is excluded (see text).
\item[(1)] Subset according to the classical Eddington ratio, $\lambda$.
\item[(2)] $\lambda$ range in logarithmic unit.
\item[(3)] Number of sources.
\item[(4)] Median black hole mass in logarithmic solar mass unit.
\item[(5)] Median Eddington ratio $\lambda $ in logarithmic unit.
\item[(6)] Median redshift of the sources.
\item[(7)] Median rest-frame 2-10 keV luminosity of the sources in
  logarithmic units of erg s$^{-1}$.
\item[(8)] Integrated rest-frame 2-10 keV flux of the stacked spectrum in logarithmic units of $10^{-12}$ erg cm$^{-2}$ s$^{-1}$.
\item[(9)] Arithmetic mean of the 2-10 keV flux per source in logarithmic units of $10^{-13}$ erg cm$^{-2}$ s$^{-1}$.
\item[(10)] Spectral slope of the stacked spectrum in the rest-frame
  2-10 keV band. $a$: When absorption is included, a steeper value
  $\alpha_{\rm E} = 0.99\pm 0.19$ is obtained with $N_{\rm H}=(0.85\pm
  0.64)\times 10^{22}$ \psqcm.
\item[(11)] Comment  on the Fe~K feature.
\end{list}
\end{center}
\end{table*}


\begin{figure*}
\centerline{\includegraphics[width=0.75\textwidth,angle=0]{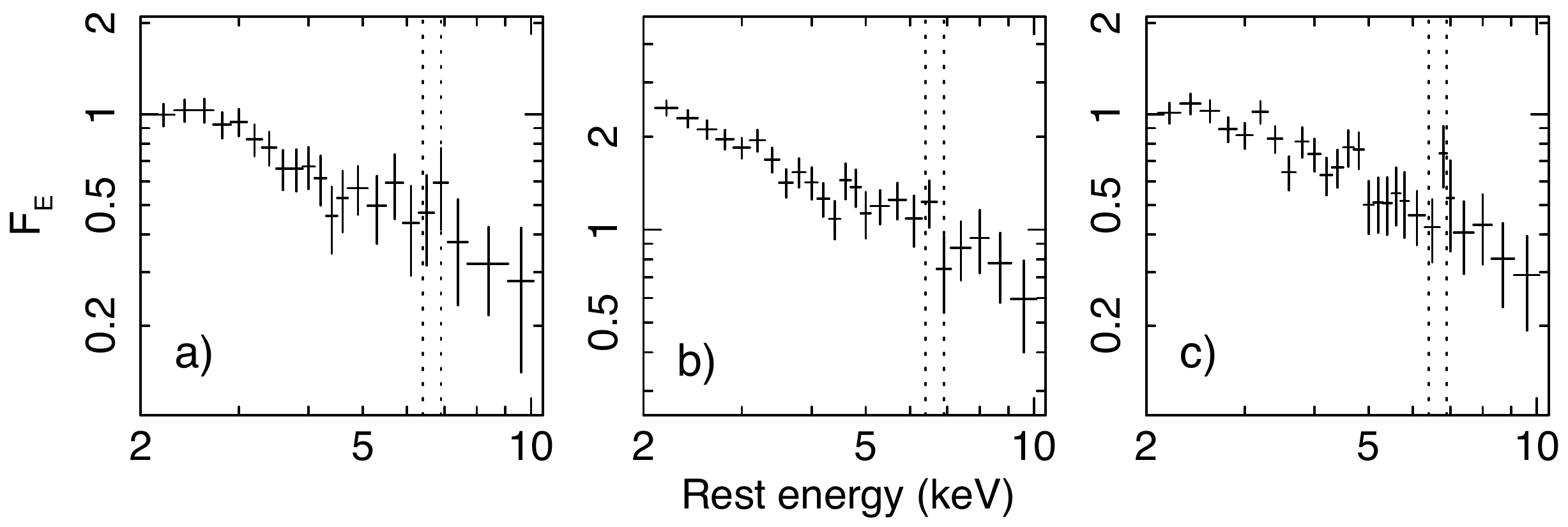}}
\caption{Stacked spectra of BLAGN (Class 1 objects) in three $\lambda
  (= L_{\rm bol}/L_{\rm Edd})$ ranges: a) log $\lambda\leq -1.3$; b)
  $-1.3< {\rm log}\thinspace\lambda \leq -0.6$ and c) log $\lambda > -0.6$ (see
  Table 2 for details). The rest energies of 6.4 keV and 6.9 keV are indicated by dotted lines. The stacked spectrum of objects with high accretion
  rates shows a strong Fe~K line at 6.9 keV.}
\end{figure*}

An investigation of the Fe~K line properties using the Class I objects
suggests that the Eddington ratio, $\lambda$, may play a role for
producing the high-ionization Fe~K, which is seen strongly in the
spectrum of Type I objects (\S~5.1 and Fig. 7). 

In the following discussion we divide the 177
objects with $\lambda $ estimates (\S 3.1.1, Fig. 5b), 
into three subsets according to $\lambda $: 1) $\lambda_{\rm L}$ (log
$\lambda \leq -1.3$),\  2) $\lambda_{\rm M}$ ($-1.3 < {\rm
  log}\thinspace\lambda\leq -0.6$),\  and 3) $\lambda_{\rm H}$ (log
$\lambda > -0.6$). The properties of the three subsets are given in
Table 2.  Fig. 10 shows the stacked spectral data from the three $\lambda $
intervals. The stacked spectrum of the objects in the high $\lambda $ range
($\lambda_{\rm H}$) shows an iron line at 6.9 keV.   Prior inspection of
individual spectra of the brightest sources in this subset shows that
a single source, XID 5230 might dominate the line feature. The EPIC pn
spectrum of this source, as observed, shows a clear line feature at an
observed energy of $3.00\pm 0.03$ keV, which corresponds to 6.95 keV
in the rest-frame of this galaxy ($z=1.32$).   Therefore this source was 
excluded from the stack (i.e. the spectrum in Fig. 10 was constructed
excluding XID 5230). However, despite the exclusion of XID 5230, the
stacked spectrum of the remainder  of the sources still shows a line feature at
$6.89^{+0.02}_{-0.03}$ keV with $EW =0.19\pm 0.10$ keV. The line flux
is comparable ($\sim$75\%) to that solely from XID 5230. A cold Fe~K line 
at 6.4 keV is not detected, with a $2\sigma $ upper limit of $EW =
0.07$ keV.

Besides the high accretion rate, the objects with $\lambda_{\rm H}$,
exhibiting the high-ionization Fe~K line, are characterized by greater
X-ray output (log $\tilde L_{2-10} = 44.52$ erg s$^{-1}$) from
relatively low-mass black holes [log $(\tilde M_{\rm
  BH}/M_{\odot})=8.1$] residing in galaxies at slightly higher
redshift ($\tilde z = 1.82$), relative to those with lower values of $\lambda$. 
These characteristics are also shared by the excluded source
XID~5230, which has $M_{\rm BH} = 8.21$, log $\lambda = -0.04$, log
$L_{2-10} = 45.68$ and $z = 1.32$. Note that 15 sources out of 39 in the
$\lambda_{\rm H}$ subset have $M_{\rm BH} < 10^8 M_{\odot}$. 
This implies that, when a stacked spectrum is made for objects
with log $(M_{\rm BH}/M_{\odot})\leq 8$, the spectrum also has a 
6.9 keV line.

While no significant line is detected, the presence of the 6.9 keV line
cannot be ruled out for the noisy $\lambda_{\rm L}$ stacked spectrum. The EW
of the $2\sigma $ upper limit is 0.35 keV. Likewise, any broad
emission would have $EW\leq 0.6$ keV. A tight upper limit for a 6.9
keV line with $EW\leq 0.08$ keV, is obtained for the $\lambda_{\rm M}$
spectrum.


There is a weak trend corresponding to an  anti-correlation between $M_{\rm BH}$ and
$\lambda $, as noted in \S~5.1.1. To verify that $\lambda $, rather
than $M_{\rm BH}$, indeed drives the spectral variations seen above,
we repeated the same exercise in a common range of $M_{\rm
  BH}$. Limiting the $M_{\rm BH}$ range to $10^{8.0} M_{\odot}$ -
$10^{8.8} M_{\odot}$, i.e., $\pm 0.4$ dex around the median value (see
Fig. 5), the numbers of objects contained in the three $\lambda $
ranges change to 31 ($\lambda_{\rm L}$), 77 ($\lambda_{\rm M}$), and
21 ($\lambda_{\rm H}$, XID 5230 excluded). The median log ($M_{\rm BH}$) 
is now found to be 8.4 for all the three $\lambda $
intervals. Spectral properties obtained from these data agree well
with those without the $M_{\rm BH}$ restriction, except for a slight
steepening of the continuum slope, $\alpha_{\rm E}=0.95\pm 0.09$,  and
an increased $EW=0.27\pm 0.14$ keV for the $\lambda_{\rm H}$
subset. This verifies that $\lambda $ is indeed a likely driver of the
variation in the Fe line characteristics found above.

\subsubsection{Spectral slope}

Previous studies have shown that the X-ray spectral slope steepens with 
increasing Eddington ratio, $\lambda $ (e.g., Shemmer et al 2006;
Porquet et al 2004). This trend is not clearly seen in our Class I
objects (Table 2). The spectrum for the $\lambda_{\rm H}$ objects might have a
curvature at lower energies which could be attributed to extra
absorption. If absorption is included, the continuum slope would
steepen to $\alpha_{\rm E} = 0.99\pm 0.19$, with $N_{\rm H} =
(0.9\pm 0.6)\times 10^{22}$ \psqcm. This slope is similar to that ($\simeq
0.95$) obtained when the $M_{\rm BH}$ restriction is applied (without
extra absorption). Even taking this steepened slope, a change in slope
$\Delta\alpha_{\rm E}\sim 0.1$ over the 1.2 dex interval in $\lambda $
between the medians of $\lambda_{\rm L}$ and $\lambda_{\rm H}$ appears
smaller than that previously reported,  e.g., in Shemmer et al (2006).  Conversely,
if an $\alpha_{\rm E}$--$\lambda $ relation is true, then the 
apparently similar slopes between the three subsets would suggest
increasing absorption with increasing $\lambda$, which would cancel out any such
steepening of the slope\footnote{A spectral curvature due to
  absorption appears in a stacked spectrum only when the absorbing columns
  are similar between objects, otherwise they effectively make a
  power-law slope flatter.}. Such absorption may be due to a partially
ionized (warm) medium so that the visibility of the optical BLRs are not
affected.

\subsection{Type II objects}


\begin{figure*}
\centerline{\includegraphics[width=0.75\textwidth,angle=0]{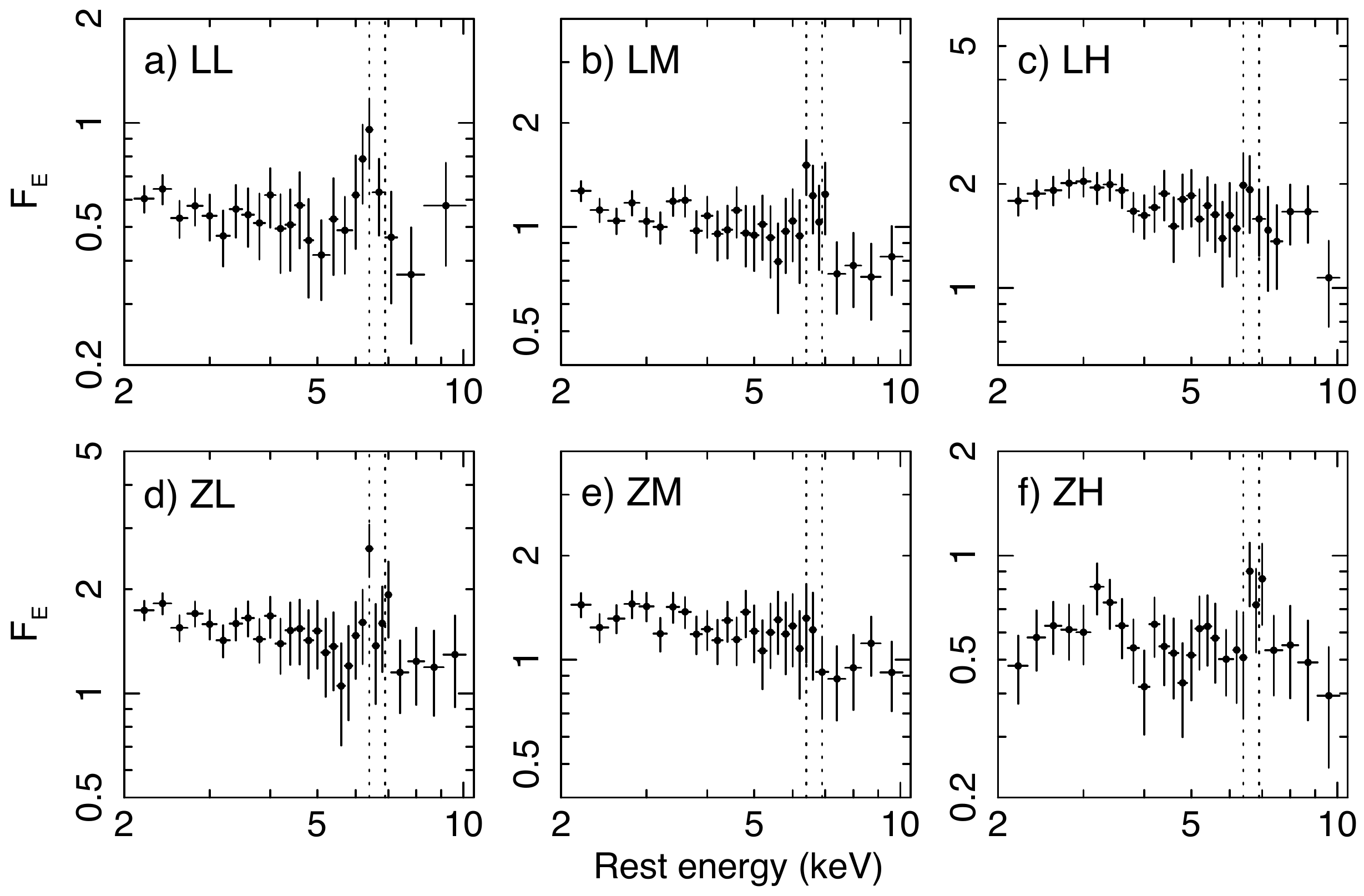}}
\caption{Stacked spectra of Type II objects. The upper three panels show luminosity slices a) LL; b) LM;
  and c) LH (Table 3). The lower three panels show redshift slices d)
  ZL; b) ZM; and c) ZH. The rest energies of 6.4 keV and 6.9 keV are indicated by dotted lines.}
\end{figure*}

The stacked spectrum of Type II objects shows a hard continuum ($\alpha
=0.22\pm 0.05$) with a strong Fe~K feature (Fig. 7). 
The Fe~K feature is resolved into two components at
$6.42^{+0.06}_{-0.07}$ keV and at $6.91^{+0.15}_{-0.03}$ keV, with a 
line intensity ratio of 1.0\thinspace :\thinspace 0.4. The two lines
are both unresolved individually and have $EW=0.15\pm 0.05$ keV and
$EW = 0.08\pm 0.06$ keV, respectively. We refer to Mainieri et al (2011)
who present the X-ray results and the host galaxy properties for a
subset of this sample, i.e. Type II QSOs with high luminosity and
significant X-ray absorption.

The spectral properties are also investigated by dividing the Type II
dataset into three ranges of the 2-10 keV luminosity and three ranges of
redshift (Fig. 11);  their basic properties are listed in Table 3. The
continuum slope remains nearly constant, $\alpha \sim 0.25\pm 0.07$, 
except  for the spectrum of the ZH ($z>2$) sample where $\alpha = 0.14\pm 0.11$. 
Interesting variations in Fe~K emission were also found, as
reported below (see Fig. 12).


\begin{figure}
\centerline{\includegraphics[width=0.48\textwidth,angle=0]{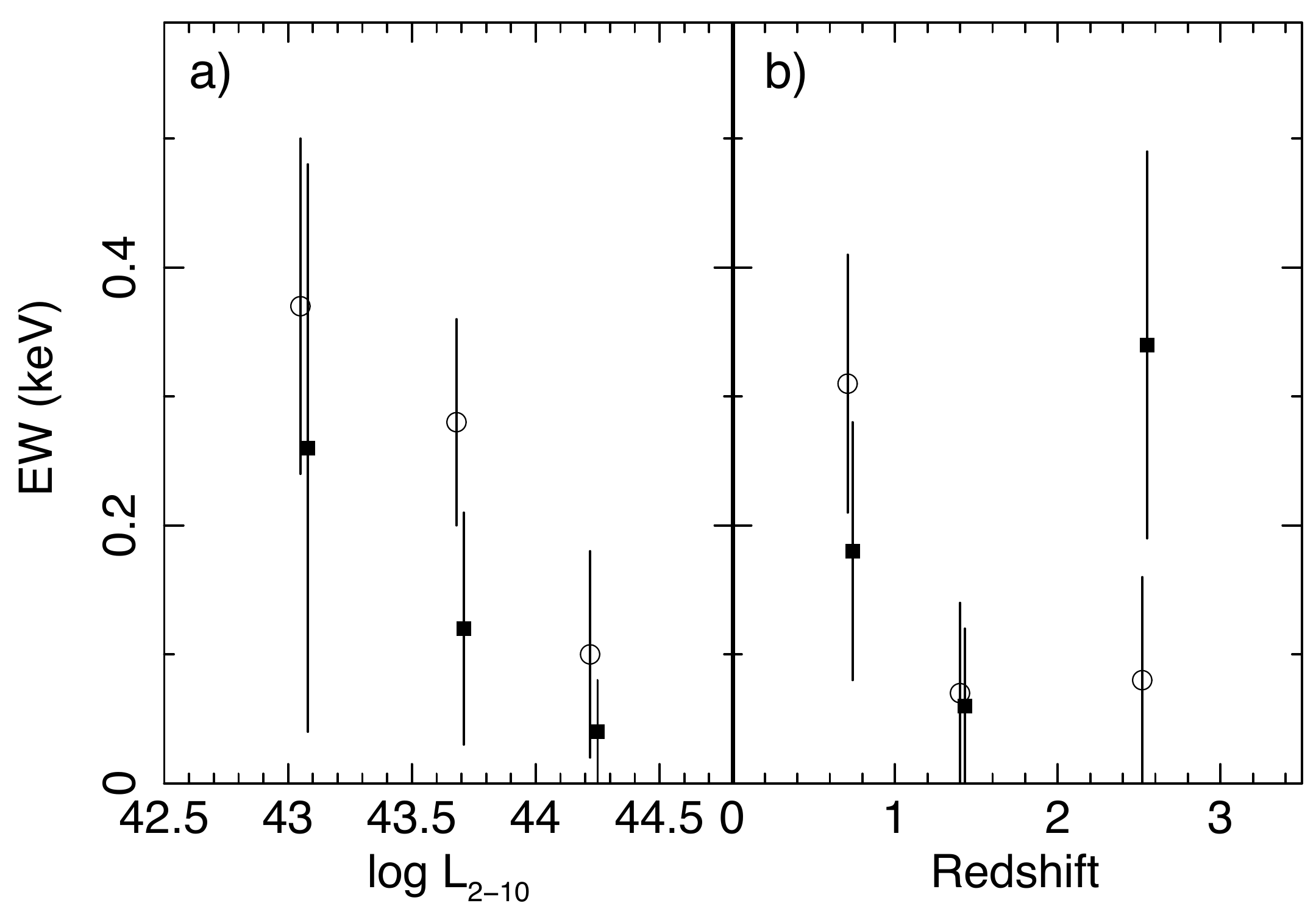}}
\caption{Plot of EW of the Fe~K line at 6.4 keV (open circles) and 6.9 keV
  (filled squares) measured from a) the three 2-10 keV luminosity bins
  or b) the three redshift bins (see Table 3). The EW is plotted against
  the median value of luminosity/$z$ in each bin. Note that the high-z
  point for the 6.9 keV line is a sum of Fe {\sc xxv} (6.70 keV) and Fe
  {\sc xxvi} (6.97 keV).}
\end{figure}

\subsubsection{Luminosity dependence}

The three luminosity intervals (LL, LM and LH) are shown in Table 3
(see also Fig. 3b) and the stacked spectra of sources in those ranges
are shown in Fig. 11a,b,c, respectively. The Fe~K feature weakens as
the 2-10 keV luminosity increases (Fig. 12a). The cold line at 6.4 keV
(open circles in Fig. 12a) is seen in all three luminosity ranges but
its EW decreases with increasing luminosity, similar to what is seen
in Type I AGN. The high-ionization line at 6.9 keV shows a similar
decreasing trend (filled squares in Fig. 12a), with a line flux ratio
of $\sim 0.7$ compared to the cold line. This trend can be compared
with the earlier result by Krumpe et al (2008).

\subsubsection{Redshift dependence}

Stacked spectra for sources in the three redshift intervals,
ZL ($z\leq 1$), ZM ($1<z\leq 2$) and ZH ($z>2$) (see Table 3 and Fig. 3b)
are shown in Fig. 11d,e,f, respectively. Only the spectrum for objects with $z\leq
1$ shows the 6.4 keV line as well as the high-ionization line at 6.85
keV. In the intermediate interval ($1<z\leq 2$), no significant line
is detected, although a weak excess is seen around 6.5 keV ($EW \leq
0.15$ keV). However, at $z>2$, a significant Fe~K feature
reappears. There are sufficient statistics to allow this feature to be
resolved into two components: Fe {\sc xxv} (6.70 keV) and Fe {\sc
  xxvi} (6.97 keV) with $EW = 0.21\pm 0.1$ keV and $EW = 0.13\pm 0.07$
keV, respectively. No 6.4 keV line is seen (the $2\sigma $ upper limit
for the  EW is 0.1 keV). 

One interpretation for the absence of a line feature in the $z = 1$-2 stacked spectrum is
that the iron atoms are predominantly in intermediate ionization
states where the fluorescence yield is low (due
to resonant trapping, e.g., Krolik \& Kallman 1987), before moving up
to higher ionization stages, as seen at higher $z$. Alternatively,
since a large number of objects in this redshift range have only
photometric redshifts (127 out of 185) any emission feature might be blurred. 
 However, the $z>2$ subset (ZH) also has a high fraction of
objects with photometric redshifts (47 out of 60), but the Fe line is
clearly detected.

As a remark related to \S~5.2.1, although $z$ and $L_{2-10}$ are,
in principle, correlated, the zones occupied by the high $L_{2-10}$ (LH in
Table 3) and the high $z$ subsets (ZH) in the $z$--$L_{\rm X}$ plane 
substantially differ from each other (see Fig. 3b
and Table 3); their median luminosities are similar but the median
redshifts are 1.76 (LH) and 2.52 (ZH), thus ZH is practically the
high-z subset of LH. We discuss the origin of the clear 
appearance of the high-ionization Fe line when the redshift range is
moved to $z > 2$ (\S 6.2 and \S 6.3).

\begin{table}
\begin{center}
\caption{Luminosity and redshift intervals of Type II objects.}
\begin{tabular}{ccccccc}
Subset & Range & $\tilde L$ & $\tilde z$ & $N$ & $F$ & $F/N$ \\
(1) & (2) & (3) & (4) & (5) & (6) & (7) \\[5pt]
& log $L_{2-10}$ &\multicolumn{5}{c}{} \\
LL & 41.00-43.35 & 43.05 & 0.61 & 162 & 0.40 & 0.25 \\
LM & 43.35-43.95 & 43.68 & 1.03 & 211 & 0.76 & 0.36 \\
LH & 43.95-45.50 & 44.22 & 1.76 & 139 & 1.30 & 0.94 \\[5pt]
& $z$ & \multicolumn{5}{c}{} \\
ZL & 0-1 & 43.31 & 0.71 & 267 & 1.14 & 0.43 \\
ZM & 1-2 & 43.84 & 1.40 & 185 & 0.91 & 0.49 \\
ZH & 2-5 & 44.30 & 2.52 & 60 & 0.43 & 0.71 \\
\end{tabular}
\begin{list}{}{}
\item[(1): ] Luminosity or redshift subset.
\item[(2): ] Range of variable, either 2-10 keV luminosity or redshift.
\item[(3): ] Median 2-10 keV luminosity.
\item[(4): ] Median redshift.
\item[(5): ] Number of sources contained in the subset.
\item[(6): ] Integrated rest-frame 2-10 keV flux in unit of $10^{-11}$
  \ergpspsqcm.
\item[(7): ] Mean rest-frame 2-10 keV flux per source in unit of
  $10^{-13}$ \ergpspsqcm.
\end{list}
\end{center}
\end{table}

\section{Discussion}

\subsection{Broad Fe emission}

Broad Fe~K emission has been found in spectra of nearby unobscured
AGN. Given that the primary X-ray production likely occurs near the 
innermost radii of the accretion disk,  and illumination of the disk
causes the line emission, relativistic broadening is considered to be
a likely explanation (e.g., Fabian et al 2000), while a complex
absorption model has been proposed to explain the redward asymmetry
(e.g., Miller, Turner \& Reeves 2009). Whatever the origin of this broad 
feature, previous investigations of Fe line profiles of nearby, bright
Seyfert galaxies found that the fraction of objects exhibiting such
broadening is $f_{\rm b}\sim 40$ \%, for which the mean EW of the broad
lines is $\bar{EW} \simeq 0.1-0.15$ keV (Nandra et al 2007; de la Calle
P\'erez et al 2010). This implies  that the mean spectrum of these
galaxies would have a broad line with $EW = 0.04-0.06$ keV. Our adopted
stacking method means that a direct comparison cannot be made, but
this value is far below our detection limit. Therefore, no detection of broad
emission is perhaps not a surprise. 

On the other hand, when the low luminosity range or intermediate
Eddington ratio are selected, possible broad line signals are seen
(sect. 5.1.1). If the broad feature with $EW \sim 0.2$ keV is real,
strong, broad Fe emission could be ubiquitous ($f_{\rm b}\simeq 1$)
among Type I AGN with $L_{2-10}\sim 10^{44}$ \ergps\ and $\lambda\sim
0.1$ at $z\sim 1.5$. This has to be tested by observations of
individual galaxies that meet those selection characteristics, and may
give impetus for a future large X-ray telescope mission. It is also in
line with the theoretical prediction made by Ballantyne (2010),
although the favourable luminosity range is lower. In nearby AGN,
objects with high $\lambda $ like MCG--6-30-15 (e.g., Tanaka et al
1995) seem to show broad Fe K emission more often than normal Seyfert
1 galaxies, which is contrary to our finding. However, in a nearby AGN
sample of FERO (de la Calle P\'erez et al 2010), no $\lambda $
dependence in the detectability of broad emission is reported.

\subsection{High-ionization Fe~K in distant AGN}


High-ionization Fe~K emission from Fe {\sc xxv} + Fe {\sc xxvi} appears
to increase in importance compared to  the cold line found at 6.4 keV in
the spectra of the XMM-COSMOS AGN, compared to
well-studied,  nearby active galaxies. The ionized gas emitting these 
lines is in such a high ionization state that other lighter elements
are expected to be almost fully stripped. These high-ionization lines
are found to be the major component of the Fe~K complex in the spectra
of Type I AGN with high Eddington ratio ($\lambda > 0.25$,
\S~5.1.2), and Type II AGN at high redshift ($z>2$,
\S~5.2.2). Here we discuss the origin of the high-ionization Fe~K
emission in the two populations as well as a possible connection between the
two.

\subsubsection{BLAGN with high Eddington ratio }

The investigation of BLAGN in \S~5.1.2 suggests that there may be a
causal link between the high-ionization Fe~K and $\lambda $. It has
been reported that PG quasars with a high $\lambda $ at low redshift,
e.g., $z<0.4$, the same range as the CAIXA sample, show
high-ionization Fe~K (Nandra et al 1996; Porquet et al 2004; Inoue,
Terashima \& Ho 2007) as well as some narrow-line Seyfert 1 galaxies
(Comastri et al 1998) that are also considered to have high $\lambda$.

With a high accretion rate, strong radiation heats the surface of the
accretion disk and probably causes the inner part of the disk to
inflate. This region is then expected to be highly ionized. The
observed 6.9 keV line is narrow however, which excludes this 
relativistic region as the origin. Even in the absence of relativistic
broadening, Fe emission produced in a highly ionized disk is
broadened by Compton scatterings within the hot layer of the disk
(e.g., Ross \& Fabian 2005). Further broadening by relativistic
effects might make such line emission difficult to detect.  

Instead,  the tenuous disk atmosphere which is likely outflowing due to
X-ray heating (e.g., Shields et al 1986) or torus winds which
evaporate off the illuminated inner surface of the obscuring torus
(Krolik \& Begelman 1986; Balsara \& Krolik 1993),  could be a likely
source of the high-ionization line. Such line emission is well
predicted in both cases (Shields et al 1986; Krolik \& Kallman
1987). 
The lack of a 6.4 keV line suggests that the covering factor of thick,
cold clouds in these objects is small, although the $2\sigma $ upper
limit is still consistent with that expected from the EW-$L_{\rm X}$
relation.


\subsubsection{High-z Type II objects:\ the local analogue and Eddington ratio}

The Type II objects at $z>2$ (ZH) show an Fe~K complex dominated by
the high-ionization lines while  a cold line at 6.4 keV is undetected.  We
note that the hard continuum spectrum ($\alpha\simeq 0.1$), which is
comparable or possibly even harder than those for the spectra in the
lower luminosity/redshift ranges, indicates that a significant
fraction of these ZH sources are strongly absorbed, yet the observed X-ray
luminosity is very large (the median log $L_{2-10} = 44.30$ \ergps; see 
Table 3). This X-ray luminosity is only 2/3 of the median value for
the other high-ionization Fe line emitters among the  Class I objects with high
$\lambda $. Combined with the EW, these two subsets are found to have
superficially comparable line luminosities for the high-ionization
Fe.   This may indicate a common origin for the line emission, and thus
the high-z Type II objects contain black holes accreting at a high
rate (or a high Eddington ratio, $\lambda $).

The quasar-like X-ray luminosity ($\geq 10^{44}$ \ergps), the hard
continuum and the high-ionization Fe~K of the high-z Type II objects
are rarely seen in active galaxies in the local Universe. One of the
few exceptions that have been known to have all of these characteristics
is the highly obscured ULIRG, IRAS F00183--7111 ($L_{\rm ir}\sim
7\times 10^{12}L_{\odot}$ at $z=0.33$). It contains a heavily obscured
AGN with  $L_{\rm X}\sim 10^{44}$ \ergps, and shows a reflection-dominated
X-ray spectrum with a strong Fe {\sc xxv} line (Nandra \& Iwasawa
2007). A remarkable characteristic of this ULIRG is the presence of
fast outflows (up to $\sim 3000$ km s$^{-1}$) probed by optical [O{\sc
  iii}]$\lambda 5007$ (Heckman, Armus \& Miley 1990),  and the mid-IR
[Ne {\sc ii}] lines (Spoon et al 2009). A small-scale radio jet
imaged with VLBI supports the idea of an AGN driven outflow (Norris et
al 2011). Although no black hole mass estimate (hence $\lambda $) is
available, a possible connection between the nuclear outflow and the
production of high-ionization Fe~K is suggested (e.g., Nandra \&
Iwasawa 2007). In local ULIRGs, molecular outflows are common and AGN
radiation pressure seems to be a favoured driving mechanism (Feruglio
et al 2010; Sturm et al 2011;  but see Chung et al 2011).

Lusso et al (2011) are able to provide robust estimates of the
bolometric luminosity estimates and stellar mass ($M_{\star}$) for 267
Type II objects, based on their SEDs. Bearing in mind that $M_{\star}$
is not for a spheroid, but for an entire galaxy, we caution that a
blind use of the scaling relation between $M_{\star}$ and $M_{\rm BH}$
is dangerous (resulting in an over-estimate of $M_{\rm BH}$ in most
cases).  Therefore, we attempted to make a crude estimate for $\lambda
$ by using the relation obtained by H\"aring \& Rix (2004). The median
value for all 267 objects is $\tilde\lambda = 0.037$. Unfortunately,
only seven objects are found in the $z>2$ range, but they all lie at
the higher end of the distribution with a median value, $\lambda =
0.68$ (practically, they are objects with higher $L_{\rm
  X}/M_{\star}$). While the caveat above, the small number of the
subsample, and the possible evolution of the scaling relation (Merloni
et al 2010; Li, Ho \& Wang 2011) make any results inconclusive, this
indicates that at least some of these objects could indeed be obscured
AGN accreting at a high Eddington ratio.

\subsection{HLIRGs at $z\geq 2$}


Out of the 60 Type II objects in the ZH sample discussed above, mid-IR
counterparts detected in the MIPS-24$\mu$m S-COSMOS
survey (Sanders et al 2007)\footnote{The catalog is publicly available
  through IRSA, http://irsa.ipac.caltech.edu/Missions/cosmos.html} have been 
identified for 27 objects (Brusa et al 2010).  With a typical
redshift of $z=2.5$ (Table 3), the detected emission corresponds to
 rest-frame emission at $\sim 7 \mu $m with a median luminosity
of log $L_{7\mu m}=45.20$ (\ergps ). This emission probably comes from
hot dust heated by an embedded AGN. When the mean SED of nearby ULIRGs
derived from the GOALS sample (U et al 2011) is assumed, log $L_{\rm
  ir}\sim 46.3$ \ergps ,  or $\sim 10^{13} L_{\odot}$, is obtained.

These powerful mid-IR sources are found to be a dominant source of the
high-ionization Fe~K in the spectrum of the ZH sources (see Fig. 11f). The stacked
spectrum of these 27 MIPS-24$\mu$m detected sources shows a Fe~K complex with
comparable intensity of Fe {\sc xxvi} and Fe {\sc xxv} (Fig. 13). The Fe {\sc
  xxvi} line accounts for virtually all the Fe {\sc xxvi} flux in the
stacked ZH spectrum. The spectrum of the other 33 sources appears to only make a  small contribution
to Fe {\sc xxv}  (see Fig. 13).

We note that several nearby (U)LIRGs, including Arp 220, with no
apparent AGN signatures, show strong Fe {\sc xxv} (Iwasawa et al 2005,
2009), which can be accounted for by thermal emission originating from
a starburst, as the X-ray to infrared luminosity ratio for those
(U)LIRGs is log $(L_{\rm X}/L_{\rm ir})\sim -4.5$ (Iwasawa et al
2009). In contrast, the ZH objects have log $(L_{\rm X}/L_{\rm
  ir})\sim -2$ and log $L_{\rm X}\sim 44.30$ (\ergps ). This X-ray
loudness indicates that an AGN is a most likely origin for the X-ray
luminosity and the high-ionization Fe~K line in the stacked ZH spectrum.


\begin{figure}
\centerline{\includegraphics[width=0.4\textwidth,angle=0]{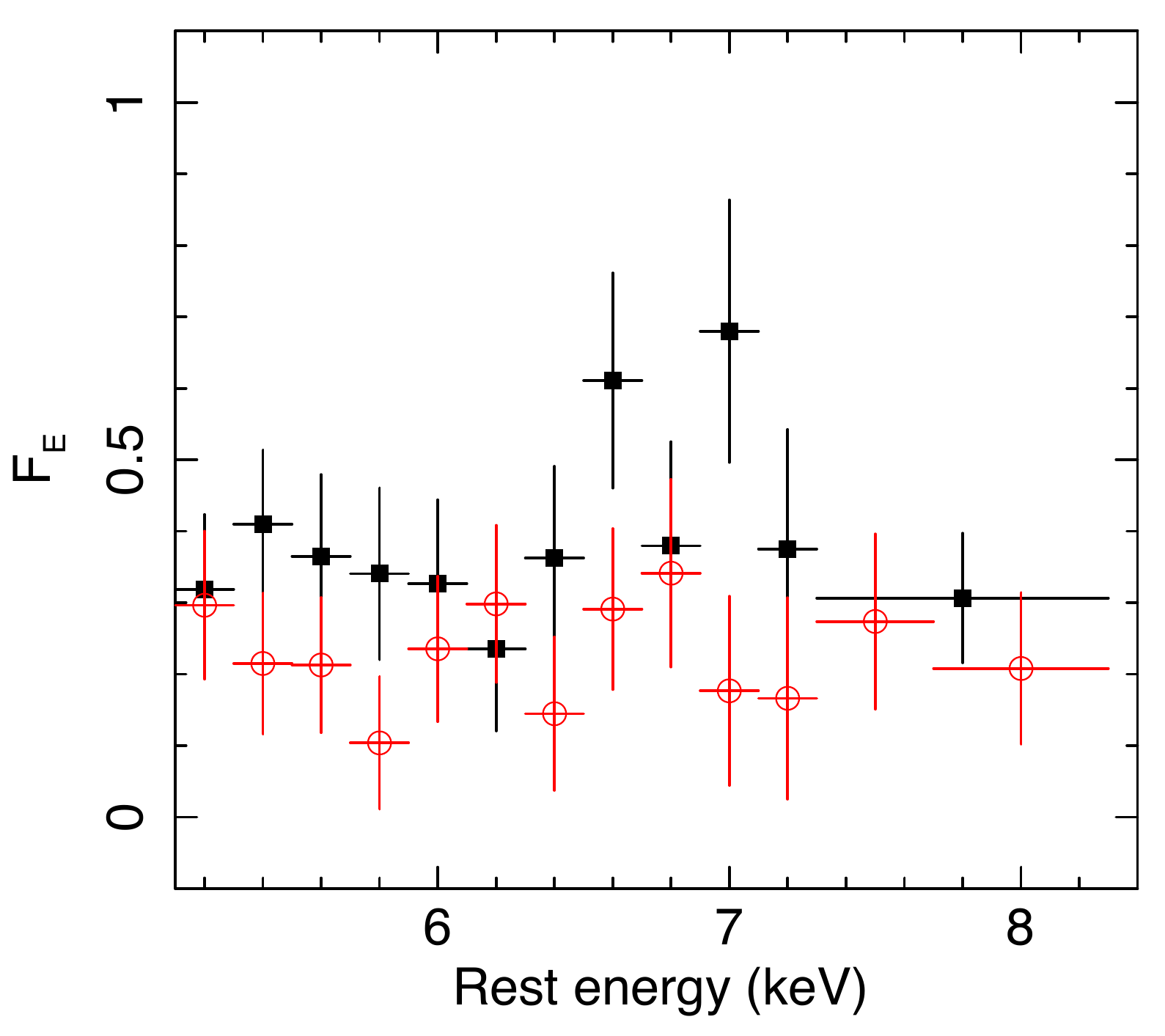}}
\caption{The Fe~K band stacked spectrum for  27 MIPS (24 $\mu $m)-detected
  objects (black solid squares) and 33 non-detected objects (red open
  circles) from the ZH subset (Table 3, Type II objects at $z>2$). The stacked 
  spectrum of the MIPS detected objects shows much stronger line emission
  and these objects are a dominant Fe~K source in  the ZH stacked spectrum of
  Fig. 11f. In particular, Fe {\sc xxvi} appears to come only from
  the MIPS-detected objects.}
\end{figure}

The number of known hyper-luminous infrared galaxies  (HLIRGs: $L_{\rm IR}\sim
10^{13}L_{\odot}$) at $z\geq 2$ have increased dramatically
after the launch of the {\it Spitzer Space Telescope}. Many of the highly
obscured HLIRGs at $z\geq 2$, selected at MIPS-24$\mu$m by Houck et al (2005),  were found to have
 IR SEDs similar to that of F00183--7111 mentioned in
\S~6.2.2., which is characterized by little PAH emission and deep
silicate absorption, suggesting a deeply buried AGN (Spoon et al
2007). Four galaxies with their infrared SEDs that  resemble  that of
IRAS~F00183--7111,  were followed up by the {\it Chandra X-ray Observatory}. Three
of these sources were detected, with $L_{2-10}\sim 10^{44}$ \ergps (Vignali et
al,  in prep). At $z\sim 2$, heavily obscured AGN are found to be
abundant in BzK-selected galaxies with MIPS-24$\mu$m detections 
using  an X-ray image stacking analysis (Daddi et al 2007; Alexander
et al 2011), although they are a much less luminous (1-2 orders of
magnitude less) population.

The MIPS-24$\mu$m detected Type II AGN at $z>2$ are, on average, powerful
infrared galaxies containing obscured AGN, as implied by the hard
continuum of the stacked spectrum ($\alpha_{\rm E}\simeq 0.1$). If
the connection between  high-ionization Fe~K and high $\lambda $ found
in BLAGN (see \S~6.2.1) is, also applicable here, it would mean
that black holes in these objects are accreting at a substantial
rate. The redshift interval where they lie coincides with the peak
period of building massive galaxies and perhaps concurrent black hole growth
(e.g, Chapman et al 2005; Hopkins et al 2006; Veilleux et al 2009).

\subsection{Compton thick AGN population at $z\simeq 0.7$}

The Type II objects in the low luminosity (LL) and  the low redshift
(ZL) intervals have  typical redshifts,  $z = 0.6-0.7$, where the major
contribution to the X-ray background is expected.  Also, the  Fe~K
features are found to be strongest in their spectra. A strong ($EW \sim
1$ keV), cold Fe line at 6.4 keV is a characteristic signature of
heavily obscured AGN. This large EW  results from the suppression
of the primary continuum due to strong absorption. Therefore, in our
stacking method of straight integration, the Fe line flux comes from
all sources while the continuum is dominated by less absorbed
sources. The EW of Fe~K in the LL and ZL stacked spectra is $EW = 0.3-0.4$ keV
(see Fig. 12). This is slightly larger than the typical value ($EW \sim
0.2$ keV) of absorbed AGN with \nH $= 10^{22}$-$10^{24}$ \psqcm\
(Fukazawa et al 2011), suggesting that there is some contribution from
Compton thick AGN to the Fe~K line. The LL and ZL sources have a typical
flux level of $10^{-14}$ \ergpspsqcm, about 2 orders of magnitude
below that of bright Compton thick AGN detected with INTEGRAL IBIS or
Swift BAT in the observed 2-10 keV band (e.g., Comastri et al
2010). Down to this flux level, no drastic change in the observed
fraction of Compton thick AGN is expected and it would remain at 10-15
\% (Gilli et al 2007; Treister, Urry \& Virani 2009 and references
therein). The observed EW (Fe~K) is consistent with this Compton thick
AGN fraction.


\begin{acknowledgements}
  This research made use of the data obtained from XMM-Newton, an ESA
  science mission with instruments and contributions directly funded
  by ESA Member States and NASA. In Italy, the XMM-COSMOS project is
  supported by ASI-INAF grants I/088/06, I/009/10/0 and
  ASI/COFIS/WP3110 I/026/07/0. In Germany the XMM-Newton project is
  supported by the Bundesministerium f\"ur Wirtshaft und
  Technologie/Deutsches Zentrum f\"ur Luft und Raumfahrt and the
  Max-Planck society. Partial support from the Italian Space
    Agency (contracs ASI-INAF ASI/INAF/I/009/10/0) is acknowledged.
  The entire COSMOS collaboration is gratefully acknowledged.
\end{acknowledgements}

\end{document}